\documentclass[fleqn,usenatbib]{mnras}
\usepackage{newtxtext,newtxmath}
\usepackage[T1]{fontenc}
\usepackage{ae,aecompl}

\usepackage{graphicx}	
\usepackage{amsmath}	
\usepackage{amssymb}	
\usepackage{subcaption} 
\usepackage{hyperref}
\usepackage{rotating}
\usepackage{afterpage}
\usepackage{multirow}  
\usepackage{filecontents} 
\usepackage{color}
\definecolor{darkgreen}{rgb}{0,0.5,0}
\definecolor{mag}{rgb}{0.79,0.08,0.48}
\definecolor{darkblue}{rgb}{0,0.2,0.8}

\title[Impact of substructure/baryon-DM transition on quads]{The impact of \(\Lambda\)CDM substructure and baryon-dark matter transition on the image positions of quad galaxy lenses}

\author[Gomer and Williams]{Matthew R. Gomer and Liliya L.R. Williams
\\
School of Physics and Astronomy, University of Minnesota, 116 Church Street SE, Minneapolis MN, 55455, USA\\
}

\pubyear{2017}

\begin{document}
\label{firstpage}
\pagerange{\pageref{firstpage}--\pageref{lastpage}}
\maketitle

\begin{abstract}
The positions of multiple images in galaxy lenses are related to the galaxy mass distribution. Smooth elliptical mass profiles were previously 
shown to be inadequate in reproducing the quad population. In this paper, we explore the deviations from such smooth elliptical mass distributions. 
Unlike most other work, we use a model-free approach based on
the relative polar image angles of quads, and their position in 3D space with respect to the Fundamental Surface of Quads.
The FSQ is defined by quads produced by elliptical lenses.
We have generated thousands of quads from synthetic populations 
of lenses with substructure consistent with \(\Lambda\)CDM simulations, and found that such perturbations are not sufficient to match 
the observed distribution of quads relative to the FSQ. The result is unchanged even when subhalo masses are increased by a factor of ten, 
and the most optimistic lensing selection bias is applied.
We then produce quads from galaxies created using two components, representing baryons and dark matter. 
The transition from the mass being dominated by baryons in inner radii to being dominated by dark matter 
in outer radii can carry with it asymmetries, which would affect relative image angles. 
We run preliminary experiments using lenses with two elliptical mass components with nonidentical 
axis ratios and position angles, perturbations from ellipticity in 
the form of nonzero Fourier coefficients \(a_4\) and \(a_6\), and artificially
offset ellipse centers as a proxy for asymmetry at image radii. We show that combination of these effects is a promising way of accounting for quad population properties. 
We conclude that the quad population provides a unique and sensitive tool for constraining detailed mass distribution in the centers of galaxies.
\end{abstract}

\begin{keywords}
gravitational lensing: strong -- galaxies: structure
\end{keywords}



\section{Introduction}
The internal structure of early-type galaxies is of great interest in the context of galaxy formation. Inner regions of galaxies are  
composed of a combination of baryonic and dark matter, forming an elliptical smooth density profile which declines with radius. 
There are different models which are used to describe the shape of such a profile, for example Sersic, NFW, and SIE (many are cataloged and described in \citet{keeton01}). 
These simple profiles do not account for several
effects which could complicate the picture, such as \(\Lambda\)CDM substructure \citep{Klypin99,moore99,springel08}, baryons that are distributed differently than dark matter, 
or line-of-sight effects in the case of lensing \citep{Jaro12,mccully16}. How big a role these effects play and how well real galaxies are described by 
simple mass distributions at 0.5-2 effective radii are the central questions this paper seeks to explore.

One of the tools that is capable of extracting information about galaxies' mass distributions is gravitational lensing \citep{blandford86}.
In the context of this paper we will specifically discuss quads, which are produced in the strong lensing regime. Five total images 
can be created, but
the central image is demagnified and difficult to detect. As a result, only four images are observed. 
The positions of the four images of quads are directly related to the distribution of mass in the main lens.  

\begin{figure}
\centering
\includegraphics[width=\linewidth]{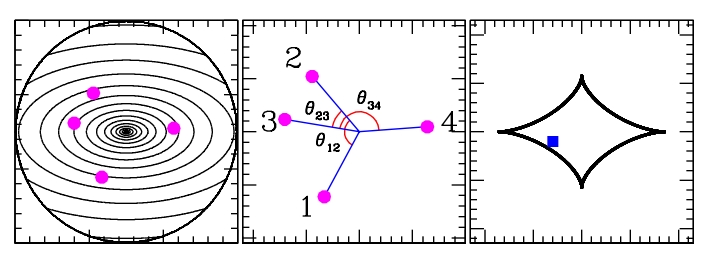}
\caption{An example quad to depict the relative image angles. The left panel shows the projected isodensity contours for an elliptical lens with the 
 images as magenta solid dots. The middle panel labels the images based on arrival order and denotes the relevant angles between the images. The right panel shows the location
 within the diamond caustic for the source from which the images arise. This figure is meant for illustration and all scales are arbitrary.}
\label{fig:exquad}
\end{figure}

In this paper we work with image polar coordinates with respect to the lens center. We use angular coordinates  because they have much less dependence on the exact radial profile 
of the galaxy while still carrying information about the level of symmetry in the lens \citep{williams08}. Like \citet{FSQ,FSQ2},
we define \(\theta_{12}\) as the angle between the first-arriving image and the second-arriving image, 
and likewise for \(\theta_{23}\) and \(\theta_{34}\), for each quad. Figure \ref{fig:exquad} shows an example of a quad with the angles labeled. The first panel shows the 
projected isodensity contours of a synthetic elliptical lens. The second panel indicates the order of arrival for each image and the definitions of relative image angles. 
The last panel shows the location inside the caustic for which a source gives rise to this quad.

\citet{FSQ} describe an empirical relationship between the relative image angles of quad images that holds true for any lens model where the 
mass distribution is symmetric about two orthogonal axes (double-mirror symmetric) and is convex at all radii 
with no wavy features (i.e. simple). 
Most parametric lens models to date fit these very general criteria, including 
psuedo-isothermal elliptical mass distributions \citep{kas93}, NFW profiles \citep{NFW96,NFW97}, de Vaucouleurs profiles \citep{deVau48},
Hernquist profiles \citep{hern90}, and others \citep{keeton01}. 

\begin{figure*}
\begin{tabular}[c]{cc}
 \begin{subfigure}[c]{.3\textwidth}
   \centering
   \includegraphics[width=\linewidth]{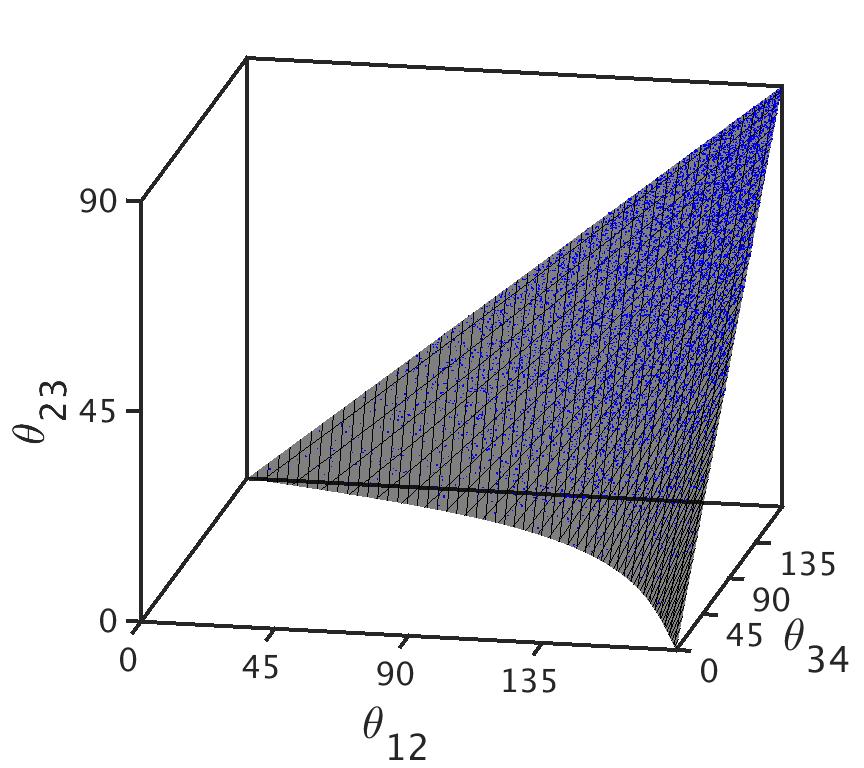}
   \label{fig:sfigfsq1}
 \end{subfigure}
 \begin{subfigure}[c]{.3\textwidth}
   \centering
   \includegraphics[width=\linewidth]{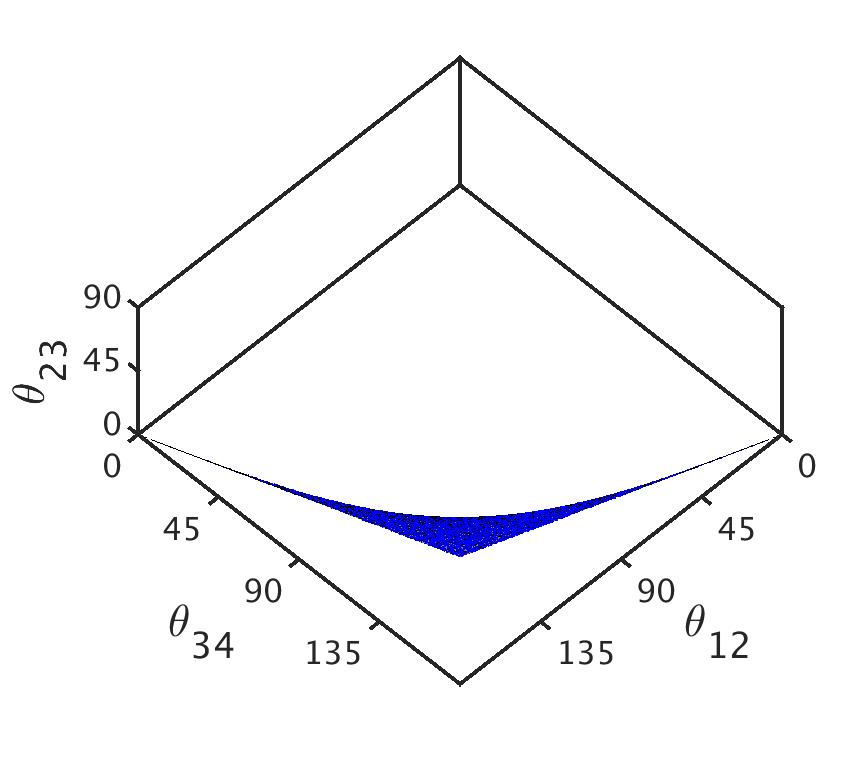}
   \label{fig:sfigfsq2}
 \end{subfigure}
\end{tabular}
 \caption{The surface shown is the Fundamental Surface of Quads (FSQ), a curved surface described via a 4th-order polynomial.
         Two different viewing angles of the FSQ are plotted in 3D, with the right panel looking ``down the barrel'' from above.
         Additionally, 10,000 quads are generated from an double-mirror symmetric elliptical lens, specifically an Einasto profile 
         with \(\alpha=0.14\) and an axis ratio of 0.82. The angles between images are plotted
         in 3D as blue points. All points lie very close to the FSQ.}
\label{fig:fsqexample}
\end{figure*}

For such lenses, plotting the relative image angles of quads in 3D results in every point lying on a surface 
called the Fundamental Surface of Quads (FSQ). We emphasize that all double-mirror symmetric lenses with elliptical isodensity contours,
independent of ellipticity or density profile, generate quads that lie very closely to the FSQ.
\footnote{There is a small caveat to the statement that all double-mirror-symmetric, convex mass distributions generate quads 
which lie on the FSQ, independent of mass profile. Different distributions within these criteria will
generate surfaces which mathematically differ from one another, albeit only slightly.  In the most drastically different models tested in \citet{FSQ}, 
deviation from the FSQ is \(\sim1.5^{\circ}\) in this 3D space, which translates to 
roughly 0.01'' on the sky. 
While some observational methods can measure to this precision or greater, the effect is negligible for our purposes.}
Figure \ref{fig:fsqexample} shows the angles of several thousand quads generated from an elliptical lens, plotted in 3D. 
The quads all lie on this well-defined surface. 

\begin{figure}
\centering
\includegraphics[width=\linewidth]{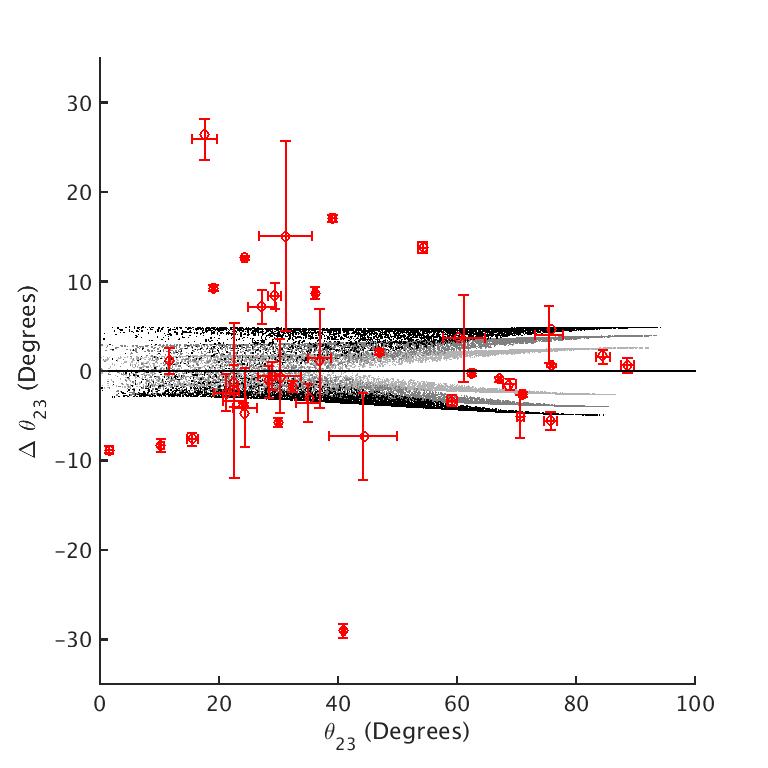}
\caption{Observed population of galaxy-scale quads from \citet{FSQ}, plotted relative to FSQ in projection (red) with error bars determined from astrometric errors in image position.
         Rather than displaying the quad angles in 3D, only the \(\theta_{23}\)-axis is shown (horizontal axis). The \(\Delta\theta_{23}\)-axis (vertical)
         depicts the difference between a quad's value for \(\theta_{23}\)
         and the value which it would need in order to lie on the FSQ. The FSQ itself is represented
         by a horizontal line at \(\Delta\theta_{23}\)=0.
         It is clear that the distribution of observed quads deviates considerably from the FSQ.
         This deviation indicates, perhaps unsurprisingly, that a purely elliptical model is too simplistic to describe the observed quad lens 
         population.          
         Additionally, quads are generated from the same lens as Figure \ref{fig:fsqexample} but with varying levels of external
         shear and plotted in grayscale. The light gray corresponds to a shear of 0.1, the darker gray a shear of 0.2, and the black a shear of 0.5. As shear increases,
         the quads lie not on the FSQ itself but on two surfaces increasingly split above and below the FSQ.}
\label{fig:shearfig}
\end{figure}

Aside from being an interesting feature which reveals some nuances in the solutions to the lens equation,
the FSQ also provides a point of comparison with observations. 
\citet{FSQ} catalog the known population of 40 quads from galaxy lenses taken from a variety of surveys.
From now on we will use a 2D projection of the FSQ (as in Figure \ref{fig:shearfig}). The horizontal axis of this projection is \(\theta_{23}\).
The vertical axis, \(\Delta\theta_{23}\), is the difference between the position of the quad and the FSQ, given \(\theta_{12}\) and \(\theta_{34}\) of the quad.
This style of plotting will be used because it allows us to visualize deviation
from the FSQ, with the FSQ itself represented as the \(\Delta\theta_{23}=0\) horizontal line.
The observed quad population with error bars from \citet{FSQ} is also plotted for comparison. 
When one plots the observed quads in comparison to the FSQ, the points do not lie exactly on the surface. This distribution of observed quads
relative to the FSQ is what we will refer to as ``observed deviations'' or ``observed discrepancy'' from the FSQ.
These deviations from the FSQ can only mean that at least some fraction of the lenses which created the observed quads are not perfect simple double-mirror 
symmetric lenses and must have some perturbations from these lens models \citep{FSQ}. Such perturbations could be anything which causes 
the mass profile to be either not double-mirror symmetric or not universally convex. In addition to external shear, candidates for such effects include 
\(\Lambda\)CDM substructure, baryonic distributions which are different from that of dark matter halos, or line-of-sight effects.

External shear in and of itself is unlikely to account for the deviations of the population of observed quads.
\citet{FSQ2} showed that the presence of external shear in an otherwise pure elliptical lens causes the population of quads to split into two surfaces-- one above and 
one below the FSQ-- instead of lying on the FSQ itself. An example is shown in Figure \ref{fig:shearfig}, which depicts quads from an elliptical lens with varying levels of shear
relative to the FSQ. 
To get deviations from the FSQ of similar magnitude as observations, rather exceptional values of shear are required.

There is an additional problem with trying to explain the FSQ deviations using shear alone.
The natural temptation is to argue that any individual observed quad which does not lie on the FSQ can be explained by 
a certain external shear tuned to describe that particular system. This would imply that the true population of quads is best represented by 
a series of split surfaces of varying heights above and below the FSQ, similar to the gray populations in Figure \ref{fig:shearfig}.
This would mean that quads exist which have large deviations from the FSQ for all values of \(\theta_{23}\). However, the observed quad population seems to 
have larger deviations from the FSQ for smaller \(\theta_{23}\) and smaller deviations for larger \(\theta_{23}\). This effect cannot be explained by external shear alone, although
external shear may be a piece of the puzzle.

This thought experiment shows the advantage of using a population of quads rather than a single quad: some 
explanations which may work for a single system cannot solve the problem for the population. To gain insight as to what kinds of lenses
are likely to be common, one must create a population of quads which is consistent with the observed population. This is the goal of the present paper.

\citet{williams08} studied the distribution of quad image angles and found statistical evidence for substructure. 
They found that simulated populations of quads from simple double-mirror symmetric elliptical lens models are unable to match the observed quad population \citep{FSQ,FSQ2}.

Though, in the context of image position angles, the exact radial density profile does not have a strong effect \citep{FSQ2}, lenses in this paper will be constructed using
Einasto profiles, which fit halos well according to simulations \citep{nav10,springel08}. 
Such profiles feature a changing logarithmic density slope, \(\gamma=-\,d\ln\rho/\,d\ln r\), which is shallower in the innermost regions and steepens
at farther radii based on the shape parameter \(\alpha\). Einasto profiles can be parameterized such that 
\begin{equation}\label{eq:einastoprof}
   \rho(r)=\rho_{-2} \exp\left\{-\frac{2}{\alpha}\left[\left(\frac{r}{r_{-2}}\right)^\alpha-1\right]\right\}  
  \end{equation}
where \(r_{-2}\) is the scale radius at which \(\gamma\)=2, corresponding to an isothermal sphere analog.
\(\rho_{-2}\) is the density at \(r_{-2}\) \citep{Einasto,springel08}.

Our goal is to construct a population of quads which is consistent with the observed discrepancy from the FSQ via a population of galaxies which 
are physically motivated. It is clear that a simple ellipsoidal smooth profile will be unable to provide enough asymmetry at the radius where images are located
to create significant deviations from the FSQ. As such, we must consider effects which cause the galaxy mass distribution to be more complicated. 
We consider two main types of asymmetries: those arising from \(\Lambda\)CDM substructure and those arising from the transition region where the mass distribution 
changes from being dominated by baryons to being dominated by dark matter. In Section \ref{sec:substructure} we will discuss the synthesis of galaxies with \(\Lambda\)CDM 
substructure and the effects of substructure on the quad distribution relative to the FSQ. It will turn out that substructure at the \(\Lambda\)CDM level will be 
insufficient to explain the FSQ deviations, so in this section we also conduct an experiment by increasing the mass of substructure by a factor of ten. 
In Section \ref{sec:biases} we will discuss the potential effects of selection biases on the quad population and the role these biases play with 
regard to the FSQ. Section \ref{sec:ellpert} discusses the effects of adding a baryon component to the mass distribution which is not identical to the dark matter. 
Finally Section \ref{sec:conclusion} summarizes these results and discusses their implications.

Throughout the paper, we use \(H_0=70\text{km s}^{-1}\text{Mpc}^{-1}\), \(\Omega_m=0.3\), and \(\Omega_{\Lambda}=0.7\).
\section{Substructure}\label{sec:substructure}
Of all phenomena potentially capable of causing deviations from the FSQ, the first candidate we explore 
is that of dark matter subhalos as predicted by \(\Lambda\)CDM simulations \citep{springel08,nav10}
and detected by multiple observations \citep{Dalal02,vegetti10,vegetti12,hezaveh16}. First we will discuss the process by which we generate lenses with such
substructure, then we will examine the effect this substructure has on the distribution of quads relative to the FSQ.

\subsection{Determining profiles for halos and subhalos}\label{ssec:math}
We synthesized lenses in 3D with properties similar to those from the Aquarius Project simulations \citep{springel08,nav10} in that they consist
of many assorted subhalos within a single main halo.
We need to be able to construct these subhalos in terms of Einasto profiles, parameterized only by the halo mass and shape parameter. 
To this end, we utilize the information gleaned about subhalo masses and distributions from \citet{springel08}, 
who present several relations which can be used to connect scale radius, central density, and total mass of a halo.
Coming from fits to simulated subhalos, the first of such relationships is between the maximum circular velocity of particles within a subhalo, \(V_{max}\),
and the total mass of that subhalo, \(M_{sub}\):
\begin{equation}\label{eq:velocity}
    M_{sub} \simeq 3.37\times10^7\left(\frac{V_{max}}{10 \text{ km s}^{-1}}\right)^{3.49} [\text{M}_{\odot}]
  \end{equation}

The maximum circular velocity  is also presented as \(V_{max}^2=11.19Gr_{-2}^2\rho_{-2}\), where \(\rho_{-2}\) is the density at the scale radius. 
Eliminating \(V_{max}\) relates \(M_{sub}\), \(r_{-2}\), and \(\rho_{-2}\), but to eliminate \(\rho_{-2}\) an additional equation is needed: the Einasto profile itself, 
specifically the enclosed mass within a given radius of such a profile, again listed in \citet{springel08}.

\begin{equation}\label{eq:enclosedmass}
    M(r,\alpha) = \frac{4\pi r_{-2}^3\rho_{-2}}{\alpha}\exp\left(\frac{3\ln \alpha + 2-\ln8}
{\alpha}\right)\gamma\left[\frac{3}{\alpha},\frac{2}{\alpha}\left(\frac{r}{r_{-2}}\right)^\alpha\right]
  \end{equation}
where \(\gamma(a,x)\) is the lower incomplete gamma function.
We would like to be able to set the Einasto profile enclosed mass equal to the subhalo mass \(M_{sub}\), but to do this we need to make an approximation.
\(M_{sub}\) is defined in \citet{springel08} as the mass enclosed within the radius such that 
the density of the particles bound to the subhalo drops below the local density at the subhalo's location. This is not strictly equivalent 
to the mass determined by integrating a profile to infinity, but it should be close enough for our purposes,
which are not so much to exactly model the Aquarius Project as much as to see if \(\Lambda\)CDM subhalos are a reasonable source for the deviations from the FSQ. 
With this in mind, we can approximate \(M_{sub} \simeq M_{\infty}\), where \(M_{\infty}=\lim_{r\to\infty} M(r,\alpha)\). This allows us to 
eliminate \(\rho_{-2}\) and find a relationship between \(r_{-2}\) and \(M_{\infty}\).
\begin{equation}\label{eq:scaleradius}
    \frac{r_{-2}}{\text{kpc}}=\frac{7.887\times10^{-4}}{A(\alpha)}\left(\frac{M_{\infty}}{\text{M}_{\odot}}\right)^{0.4269}
  \end{equation}
where
\begin{equation*}
    A(\alpha) = \frac{1}{\alpha}\exp\left(\frac{3\ln \alpha + 2-\ln8}{\alpha}\right)\Gamma\left[\frac{3}{\alpha}\right]
  \end{equation*}
with \(\Gamma(x)\) being the complete gamma function. \(A(\alpha) \sim 10\) for realistic values of \(\alpha\). 
The next step is to project this Einasto profile into 2D for lensing. \citet{dhar10} showed that the 2D projected 
Einasto profile can be analytically approximated as 
\begin{flalign}\label{eq:2deinasto}
    \Sigma(X,\alpha)=\frac{\Sigma_0}{\Gamma[n+1]}\bigg\{n\Gamma\left[n,b(\zeta_2 X)^{\frac{1}{n}}\right]\\
+\frac{b^n}{2}X^{\left(1-\frac{1}{2n}\right)}\gamma\left[\frac{1}{2},X^{\frac{1}{n}}\right]e^{-bX^{\frac{1}{n}}}\nonumber
-&\delta b^n X e^{-b\left(\sqrt{1+\epsilon^2}X\right)^{\frac{1}{n}}}\bigg\}
  \end{flalign}
where \(\Gamma(x,a)\) is the upper incomplete gamma function and \(X=\frac{R}{r_{-2}}\),
with \(R\) now being the 2D projected radius from the center and \(r_{-2}\) being the same 3D scale radius as before. 
The central surface mass density at \(R=0\) is \(\Sigma_0\), while \(n=\frac{1}{\alpha}\) and \(b=2n\); \(\delta\),\(\epsilon\), and 
\(\zeta_2\) are functions of \(n\) and \(X\).
To calculate the mass within R one needs to integrate:
\begin{equation}\label{eq:einastomass}
    M(R,\alpha)= \iint \Sigma(R,\alpha)R\,dR\,d\theta
=2\pi \Sigma_0 r_{-2}^2\int_0^{\frac{R}{r_{-2}}}X s(X,\alpha) \,dX
  \end{equation}
where \(s(X,\alpha)=\Sigma(X,\alpha)/\Sigma_0\). Taking R to infinity and numerically integrating, 
we can combine Equation \ref{eq:einastomass} with Equation \ref{eq:scaleradius}, 
resulting in a relationship between total mass and central density
\begin{equation}
\begin{split}
    \frac{\Sigma_0}{\text{M}_\odot\text{kpc}^{-2}}&= 2.559\times10^5 \frac{A(\alpha)^2}{\int_0^{\infty} Xs(X,\alpha)\,dX} 
\left(\frac{M_{\infty}}{\text{M}_{\odot}}\right)^{0.1461}\\
&=2.559\times10^5 B(\alpha)\left(\frac{M_{\infty}}{\text{M}_{\odot}}\right)^{0.1461}
\end{split}
\end{equation}
where all of the \(\alpha-\)dependence is absorbed into a function \(B(\alpha)\sim10^2-10^3\). With this, we can readily convert a desired mass for a halo into a 
corresponding 2D central density and Einasto scale radius. 

Before continuing, it is prudent to check that the approximations above yield results that are reasonable. Specifically, we should
confirm that these formulas, which are derived in part from information about the subhalos, can reasonably accurately 
describe the main halo as well.
We can test this by inserting values quoted in the Aq-A-1 simulation, the highest resolution simulation discussed in \citet{springel08}.
\(M_{50}\), the mass enclosed within the radius at which the enclosed density is 50\(\times\) the average density of the universe, 
is quoted for Aq-A-1 as \(2.523\times10^{12}\text{M}_{\odot}\). The shape parameter is quoted in \citet{nav10} as \(\alpha=0.170\). Similar to 
above, one can approximate \(M_{50} \simeq M_{\infty}\) and plug this into Equation \ref{eq:scaleradius} to yield a scale radius of  12.3 kpc, 
while \citet{springel08} report a scale radius of 13.0 kpc, which gives an idea of the level of error in the above approximations. 

Now that the relationships between \(\alpha\), \(\Sigma_0\), and \(M\) have been established, a small deviation is made from the Aquarius Project, 
in that the shape parameter of the main lens is 
changed from \(\alpha=0.17\) to \(\alpha=0.14\), reducing the scale radius to 8.5 kpc. The reasoning for this is as follows.
Since the goal is to compare simulated lenses which have only a single mass component with observations which have both baryons and dark matter,
it is necessary to alter the shape of the single-component profile to be less like dark-matter-only simulations and more like observations.
\citet{slacs3} have measured the average 3D logarithmic density slope within the Einstein radius, 
\(\langle \gamma'_{3D}\rangle\ =\langle -\,d\log\rho/\,d\log r \rangle\ = 2.0^{+0.02}_{-0.03}\). 
The slope of 2.0 indicates that the effects of baryons and dark matter have ``conspired'' to make the density slope approximately ``isothermal.''
\footnote{``Isothermal'' refers to the fact that the slope is equivalent to that of an isothermal sphere, but the quotation marks are used because
the term in this context makes no claim as to the dynamics of the system, such as whether or not the system is actually at the same temperature throughout.}
The authors model the profile as a power law, meaning the average logarithmic density slope \(\langle \gamma'_{3D}\rangle\) is the same as 
the local slope at the Einstein radius, \(\gamma'_{3D}\rvert_{R_E}\). 
It is reasonable to assume that the 2D logarithmic density slope \(\gamma'_{2D}=-\,d\log\Sigma/\,d\log R\approx \gamma'_{3D}-1\approx 1\),
although due to projection effects and a changing slope, the value at the Einstein radius would not be exactly \(\gamma'_{3D}-1\).
Without changing the total mass of the halo, altering \(\alpha\) to 0.14 makes the mass profile more centrally concentrated 
than the Aquarius simulations, which increases the size of the Einstein radius, and steepens the local slope at that 
radius from \(\gamma'_{2D}=0.73\) to 0.91. Since this is closer to the target of \(\approx 1\), 
it is considered more realistic.

\subsection{Subhalo mass function and distribution within main halo}
In the Aquarius simulations, subhalos are fit by several different shape parameters with slightly
higher values than the main halo shape parameters, roughly ranging from \(\alpha_{sub}=\)0.15 to 0.21. 
In this paper \(\alpha_{sub}\) is set to 0.18 for all subhalos, similar to what is done in \citet{springel08}, which
the authors argue is roughly equivalent to having \(\alpha_{sub}\) in the range of \(0.16-0.20\). 

For substructure to be analogous to that in the Aquarius simulations, subhalos must follow a mass function:
\begin{equation}
    \frac{dN}{dM} = a_0 \left(\frac{M}{m_0}\right)^n
  \end{equation}
with \(n=-1.9\) and \(a_0=8.21 \times 10^7/M_{50}\) where \(m_0=10^{-5}M_{50}\). \(M_{50}\) refers to the main halo and will again be approximated as \(M_{\infty}\).  
Subhalo populations analogous to the Aq-A-1 simulation run by \citet{springel08} are constructed using \(M_{50}=2.5 \times 10^{12}\text{M}_{\odot}\)
The fractional mass in subhalos \(f_{sub}\) is of order 0.1, so a population of subhalo masses is synthesized to have a cumulative mass of 
\(2.5 \times 10^{11}\text{M}_{\odot}\). Optimally, a population would include subhalo masses all the way down to the free-streaming limit of dark matter \citep{springel08}, 
but this is not feasible since the inclusion of lower mass subhalos requires many more subhalos themselves, and therefore is computationally
expensive. Additionally, smaller halos are less important because they are less likely to produce the densities or shears necessary to affect the position of images.
With this in mind, each synthesized main halo contains a population of 133,400 subhalos with masses ranging from \(10^{5}\text{M}_{\odot}\) to \(10^{10}\text{M}_{\odot}\),
distributed in 3D. 

In our lensing simulations themselves, however, very few of these numerous subhalos are actually included. This is because we are interested only in the regions
which are likely to produce quads, which limits us to the central region of the galaxy. Only the subhalos which are positioned along the line of sight and near 
the center in projection will be relevant. The 3D distribution of subhalos in space is produced by an Einasto profile with \(\alpha=0.678\) and 
\(r_{-2}=199\)kpc, consistent again with \citet{springel08}. Subhalos which are farther than 17 kpc away from the line-of-sight axis-- a somewhat arbitrary 
value chosen to be slightly larger than the window of the 2D simulations-- are considered too far from the center and are omitted. 
This typically leaves \(900-1000\) subhalos remaining to be included in the lens itself.

%

To check the effect of halos outside the simulation window, one can estimate the amount by which the image positions, and therefore angles, would change.
We estimate this by treating the subhalo as a point mass, and placing it at two window radii. The largest possible mass for such a subhalo would be \(10^{10}\text{M}_{\odot}\).
We then calculate the deflection angle for a single image at the Einstein radius. For our lens and source redshifts, this corresponds to a deflection of 0.05 kpc = 0.007''.
To have the most extreme effect on releative image angles, we imagine the deflector is on the x-axis and the image is on the y-axis, so that the deflection is nearly completely in the 
azimuthal direction with respect to the center of the lens. It turns out that the polar image angle is altered by \(\sim1^{\circ}\). Additionally, since image angles are defined with respect to one 
another, there could be up to a factor of 2 additional effect if another image lies opposite the first (at \(y\simeq-R_E\)). This means that if a subhalo were placed outside the window
in such a way to optimally alter our image angles, it could only change angles by \(\sim2^{\circ}\), which is significantly less than the level of FSQ deviation necessary to match observations 
(\(\sim10^{\circ}\)). We therefore choose to simply ignore the effects of subhalos outside of the window radius.

\subsection{Comparison of simulations with observations}
\begin{figure}
\includegraphics[width=\linewidth]{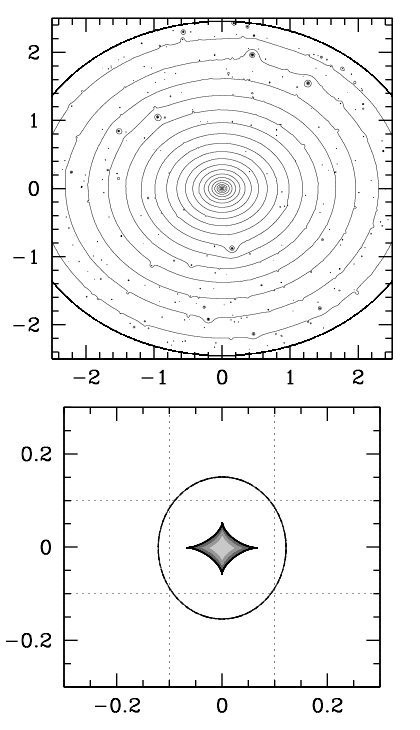}
\centering
\caption{Density contours and caustic for a lens with \(\Lambda\)CDM substructure. This lens was chosen as an example because it has larger
   perturbations due to substructure relative to some of the other realizations.
   Scale is in arcseconds. At z=0.6, 1 arcsecond corresponds to 6.7 kpc. Typical quad images are at \(\simeq2.5\)kpc from the center.
   Source positions within the diamond caustic are shaded according to resulting \(\theta_{23}\) values, with
   darker gray indicating an angle less than 40 degrees, lighter gray indicating greater than 60 degrees, and gray indicating the intermediate angles. 
   Quads more similar to ``Einstein crosses'' come from sources in the lighter gray central region. 
   10,000 quads are created from sources within the diamond caustic.}
\label{fig:cdmcontour}
\end{figure}

Each main lens is identical, with \(\alpha=0.14\) and \(M_{\infty}=2.5\times 10^{12} \text{M}_{\odot}\), corresponding to \(r_{-2}=8.5\) kpc.
The lenses are elliptical, with an axis ratio of \(q=0.82\). The critical lensing density is set to \(\Sigma_{crit}=1.85\times10^{9} \text{M}_{\odot} \text{kpc}^{-2}\),
corresponding to the source being at redshift \(z=3.0\) and the lens being at \(z=0.6\), making the Einstein radius \(R_E=2.5\) kpc.
For now lenses have zero external shear. Stacked on top of the main halos are \(\sim 900\) subhalos 
that happen to lie along the line of sight within the virial radius.
Different lenses are created via different random seeds for the subhalos.
An example lens is depicted in Figure~\ref{fig:cdmcontour}.

\begin{figure}
\includegraphics[width=\linewidth]{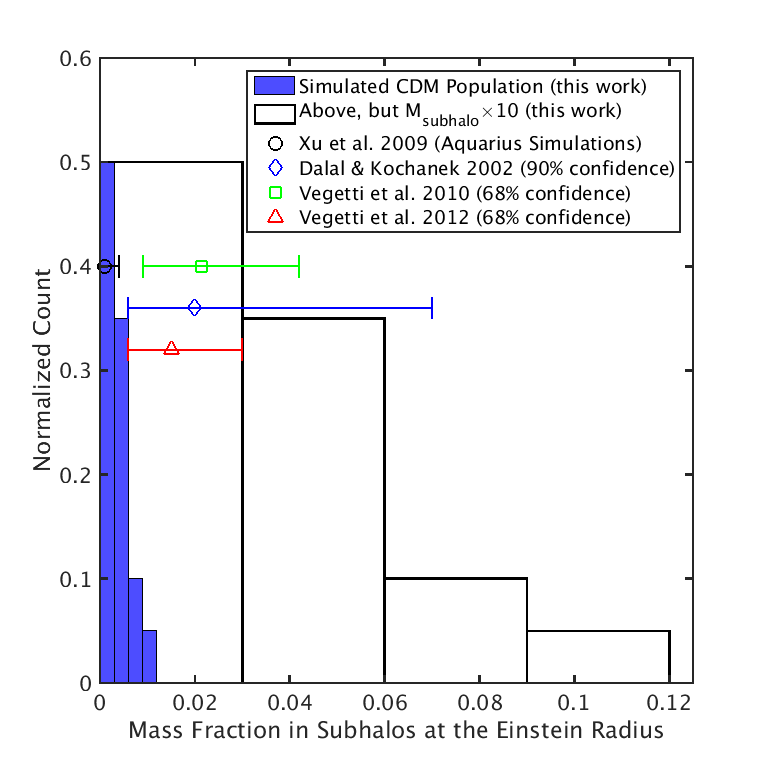}
\centering
\caption{The distribution of subhalo mass fractions localized near the Einstein radius for the population of 20 lenses is depicted as a solid blue histogram.
         The same 20 lenses, but with their substructure mass amplified by a factor of 10, have the mass fraction distribution shown as the white open histogram. 
         Our simulated \(\Lambda\)CDM lenses are consistent with 
         the values for the Aquarius simulations from \citet{xu2009} (black circle), but are on the edge of the 68\% C.L. for the observations of
         \citet{vegetti10} (green square) and \citet{vegetti12} (red triangle) and the 90\% C.L. for the observations of \citet{Dalal02} (blue diamond). 
         Vertical positions for these values are arbitrary.
         This is not necessarily inconsistent, but it does provide some inspiration for our \(10\times\Lambda\)CDM test in Section \ref{ssec:10xCDM},
         where substructure produces mass fractions closer to these observations, similar to the white histogram.}
\label{fig:massfrachist}
\end{figure}

Near the Einstein radius, our simulated lenses have projected mass fractions \(\langle f \rangle\) 
ranging from 0.1\% to 1.0\% in subhalos.
This is consistent with the Aquarius halos, which have \(\langle f \rangle = 0.1^{+0.3}_{-0.1}\%\) \citep{springel08,xu2009,vegetti12}, shown in
Figure \ref{fig:massfrachist}. The number of halos in each mass bin is consistent with the Aquarius simulations \citep{xu2015}.

It is worth noting that recent observations have detected this type
of substructure in real lenses and have inferred mass fractions which are higher than \(\Lambda\)CDM simulations. 
In a study of seven lens systems, \citet{Dalal02} deduced a local mass fraction of subhalos at the image radius between 0.6\% and 7\% with a 90\% confidence level.
\citet{vegetti10} found a dark substructure in the
lens SDSSJ0946+1006, which is one of the 40 quads in \citet{FSQ}. For this galaxy, \citet{vegetti10} infer
\(\langle{f}\rangle=2.15^{+2.05}_{-1.25}\%\) at the Einstein radius when assuming \(n=-1.9\pm0.1\). 
When comparing their value with simulations, they find a likelihood of 0.51, which 
is consistent given their sole detection. \citet{vegetti12} found a dark satellite in the JVAS B1938+666 system implying 
an average subhalo mass fraction \(\langle{f}\rangle=3.3^{+3.6}_{-1.8}\%\) within the Einstein radius and a different mass function slope \(n=-1.1^{+0.4}_{-0.6}\). Had they 
assumed \(n=-1.9\pm0.1\) they would be closer to the Aquarius simulated mass fraction with their result of \(\langle{f}\rangle=1.5^{+1.5}_{-0.9}\%\), 
arguing that the remaining discrepancy is due to the fact that their galaxy is at a different redshift than those in the Aquarius Project, which are at z=0.
These findings are compared visually with our simulated lens population in Figure \ref{fig:massfrachist}.
Finally, \citet{hezaveh16} report the detection of a dark subhalo in the SDP.81 system. They claim their mass function is consistent with simulations, based on their
one detected subhalo at \(10^9\text{M}_{\odot}\) and upper limits established at lower masses. It remains to be seen whether the minor tension between
some observations and simulations indicates a problem with theory or is simply a result of having a small number of observations.

More recent simulations which include
baryon effects have been done by \citet{fiacconi16}, which found that local projected mass fractions at the Einstein radius for \(\sim10^{13}\text{M}_{\odot}\) halos
can exceed 2\% if the lens is at a redshift of 0.7. Since the halos are both more massive and at a higher redshift than the Aquarius halo, they are 
dynamically younger, as their subhalos are accreted more recently, and therefore clumpier, with higher mass fractions.

Our lenses are meant to represent the population of observed galaxies, so it is useful to compare them with typical quad-producing galaxies. 
Unlike the real galaxy population, these synthetic lenses represent a population of galaxies which would 
all have the same mass, redshift, and profile shape. This may seem like an oversimplification,
but note that these parameters would not cause any asymmetries in the lenses. Similar galaxies at different redshifts, for example, would change
the critical densities necessary for lensing and therefore the radial positions of the images, but, since this affects all parts of the lens equally, the double-mirror 
symmetry of the lens remains unaltered. Since the double-mirror symmetry is unchanged, the deviations from the FSQ will be no different \citep{FSQ2}. This further justifies
the assumptions about the mass profile of the lenses made in Section \ref{ssec:math}. 

The ellipticity and external shear for each galaxy are also held constant for now, with an axis ratio of 0.82 and no external shear. 
Unlike the parameters above, these properties can cause deviations in the FSQ, provided the shear is at an oblique angle with
respect to the ellipse major axis. In the terminology of \citet{FSQ2}, these parameters create Type II lenses, in that they break the double-mirror symmetry only once,
as opposed to substructured lenses (Type III), which have no remaining symmetries.
The reason these parameters are held constant for now is not because they have no effect on image angles, but that we seek to isolate the effect of substructure. 
In Section \ref{ssec:shear} we will relax the restriction that ellipticity and shear be so constrained.

\begin{figure}
\begin{subfigure}{.46\textwidth}
  \centering
  \includegraphics[width=\linewidth]{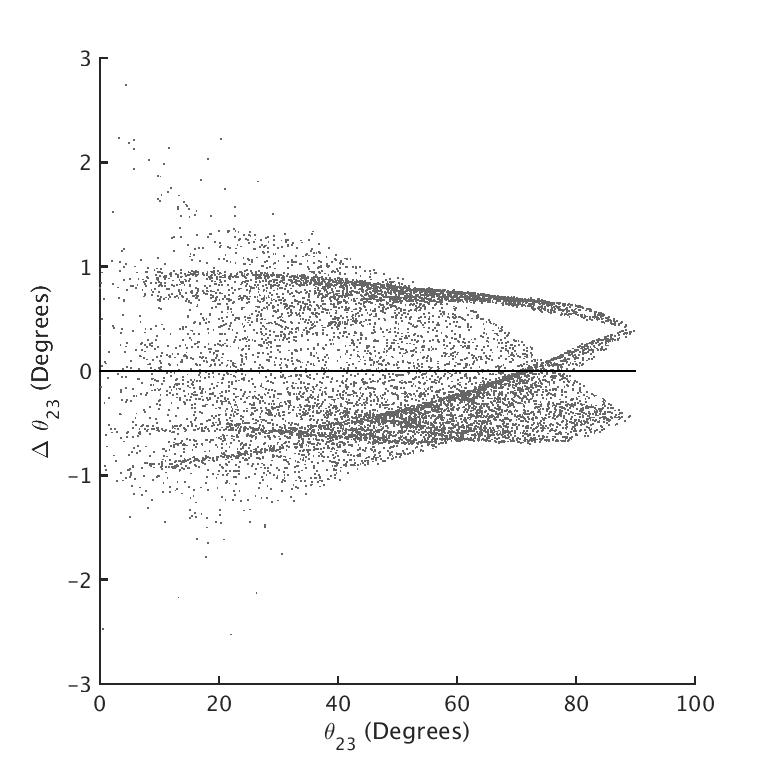}
  \label{fig:sfigsing1cdm}
\end{subfigure}
\begin{subfigure}{.46\textwidth}
  \centering
  \includegraphics[width=\linewidth]{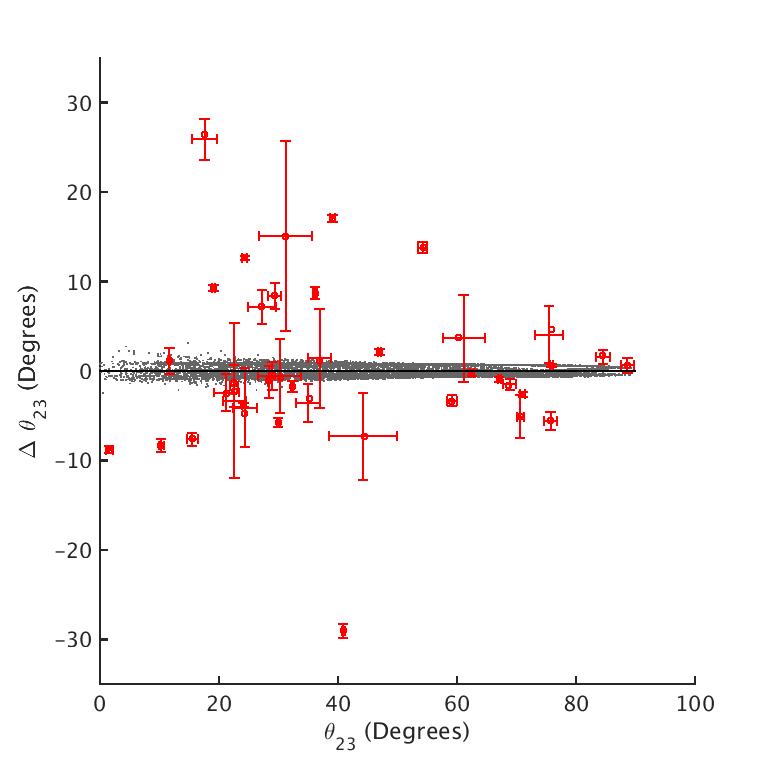}
  \label{fig:sfigsing1cdmobs}
\end{subfigure}
\caption{The top panel shows the deviations of the quad relative image angles from the FSQ
      in a fashion similar to Figure \ref{fig:shearfig}, but for the single lens with \(\Lambda\)CDM subhalo perturbers shown in Figure \ref{fig:cdmcontour}. 
      Each point represents a quad. For a purely elliptical lens,
      the deviations would be very nearly zero. The bottom panel is the same except the vertical scale is increased and 
      observed quads are included as red circles. It is immediately apparent that the simulated deviations, 
      while nonzero, are insufficient to explain the observed deviations.
      While not depicted here, all other attempted random realizations of substructure positions have similar results.}    
\label{fig:1cdmfsq}
\end{figure}

From the example lens we created 10,000 quads and determined their deviations from the FSQ. The FSQ is described by a polynomial fit explicitly shown in
\citet{FSQ} which expresses \(\theta_{23}\) as a function of \(\theta_{12}\) and \(\theta_{34}\). For each quad, the difference between the FSQ-predicted
\(\theta_{23}\) and the actual value for each quad is used as the measure of the deviation. In Figure \ref{fig:1cdmfsq}, the resulting deviations from the FSQ
for the example lens are shown. The population of 40 observed quads cataloged in \citet{FSQ} is plotted alongside our simulated quads in the bottom panel of Figure \ref{fig:1cdmfsq}.
It is evident by eye that the scale of deviations provided by \(\Lambda\)CDM substructure is simply too small to account for observations. 
\subsection{Testing larger subhalos than CDM}\label{ssec:10xCDM}
\begin{figure*}
 \centering
  \begin{tabular}[c]{cc}
   \begin{subfigure}[c]{0.3\textwidth}
    \includegraphics[width=\linewidth]{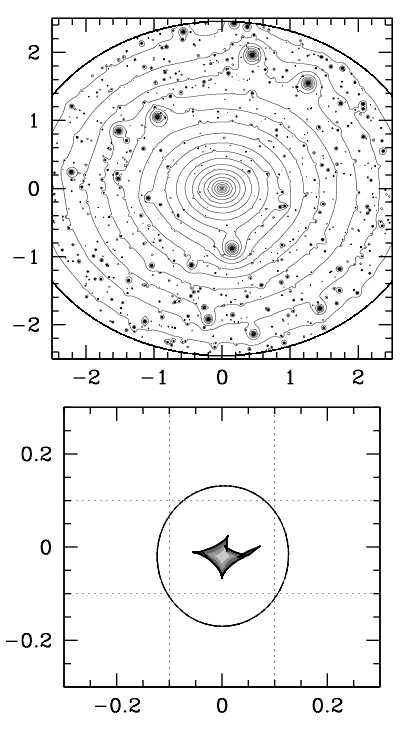}
    \label{fig:sfig10xseed18}
   \end{subfigure}&
  \begin{subfigure}[c]{0.3\textwidth}
   \includegraphics[width=\linewidth]{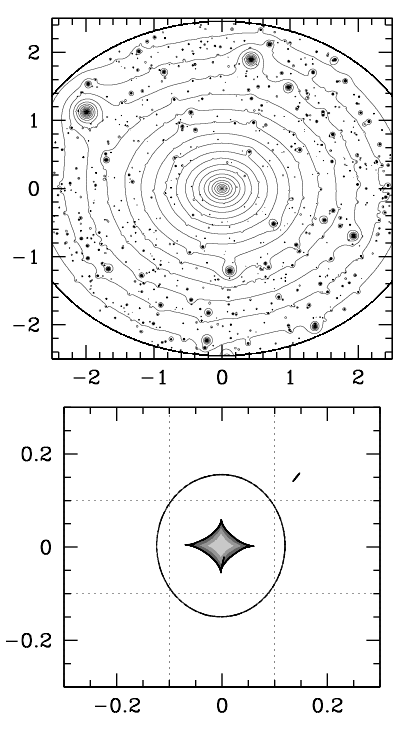}
   \label{fig:sfig10xseed05}
  \end{subfigure}
 \end{tabular}
\caption{The left lens is the same as in Figure \ref{fig:cdmcontour} except with the mass normalization for subhalos multiplied by a factor of 10.
   Note the disturbed caustic. The \(10\times\)CDM-population
   consists of 200 lenses similar to this one, with different subhalos generated drawn from the full 3D population. 
   The lens on the right is also \(10\times\)CDM, with this particular realization generating a more tame caustic. Again, 1 arcsecond corresponds to 6.7 kpc. }
\label{fig:10xCDMcontour}
\end{figure*}

Because substructure at the scale of \(\Lambda\)CDM simulations is too weak of a perturbation to create the observed deviations from the FSQ, we now
increase the mass of the subhalos to see what the effects would be. This idea is motivated in part by the potential tension mentioned above between the subhalo mass
fractions of simulations and observations of \citet{vegetti10,vegetti12,hezaveh16}, where observations may hint at more mass in subhalos than what \(\Lambda\)CDM simulations predict.
The logic is essentially that if subhalos are increased in mass, they may be able to reproduce the deviations from the FSQ. 
That would hint that perhaps the predictions from \(\Lambda\)CDM simulations do not create sufficiently large subhalos to match reality. 
On the other hand, if even larger subhalos are still unable to recreate the FSQ deviations, that would indicate that \(\Lambda\)CDM subhalos are undeniably not responsible for the 
observed deviations from the FSQ. With this in mind, we dial up the normalization for subhalos by a factor of ten above the \(\Lambda\)CDM prediction and re-run the experiment. 
The same example lens with now \(10\times\) the subhalo mass as \(\Lambda\)CDM is shown in Figure \ref{fig:10xCDMcontour}, in addition to a second lens with a different seed.

A complication that arises for some subhalo configurations is that since subhalos are a factor of ten larger than those produced 
with \(\Lambda\)CDM, they are more likely to have central densities high enough
to be their own strong lenses. We are not interested in
these types of scenarios because the images are so close together that they are unlikely to be resolved in observations.
To prevent situations like these two steps are taken. First, if the last-arriving image (central, 5th image) is more than 0.5\(R_E\)
from the lens center the quad is not included. Second, when comparing with observations, we will use quads only within a selection window which omits the
quads with \(\theta_{23}<10^{\circ}\), where these anomalies are most likely to reside. 
This selection also omits one observed quad. 

\begin{figure*}
\begin{subfigure}{.33\textwidth}
  \centering
  \includegraphics[width=\linewidth]{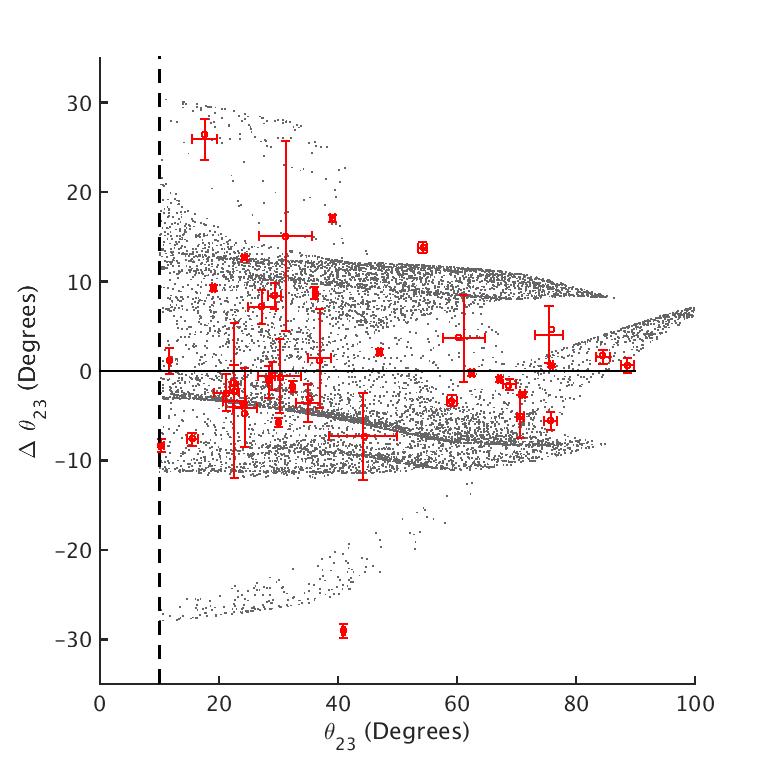}
  \label{fig:sfigsing10cdmobs}
\end{subfigure}
\begin{subfigure}{.33\textwidth}
  \centering
  \includegraphics[width=\linewidth]{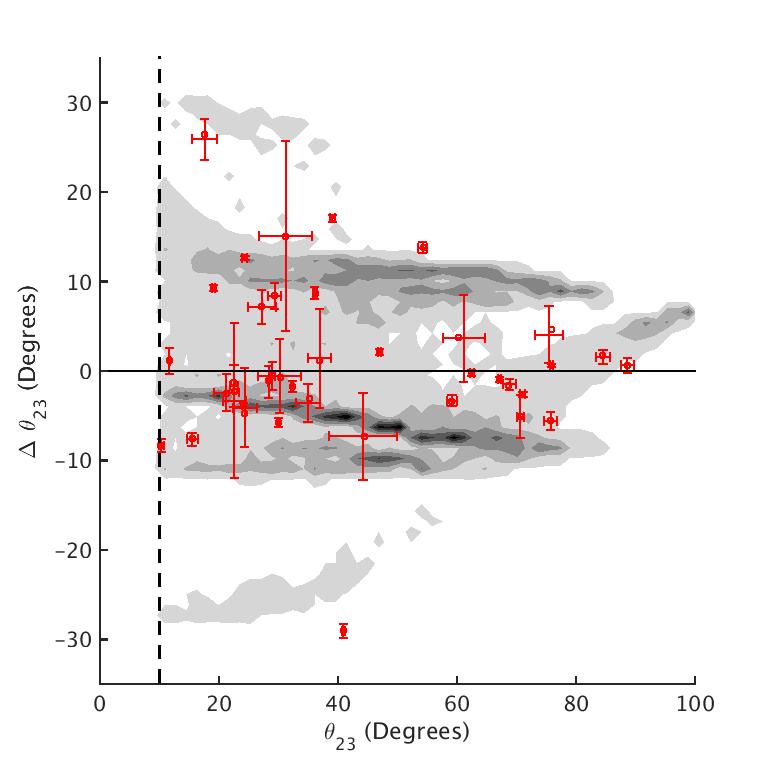} 
  \label{fig:sfigsing10cdmcontour}
\end{subfigure}
\begin{subfigure}{0.33\textwidth}
  \centering
  \includegraphics[width=\linewidth]{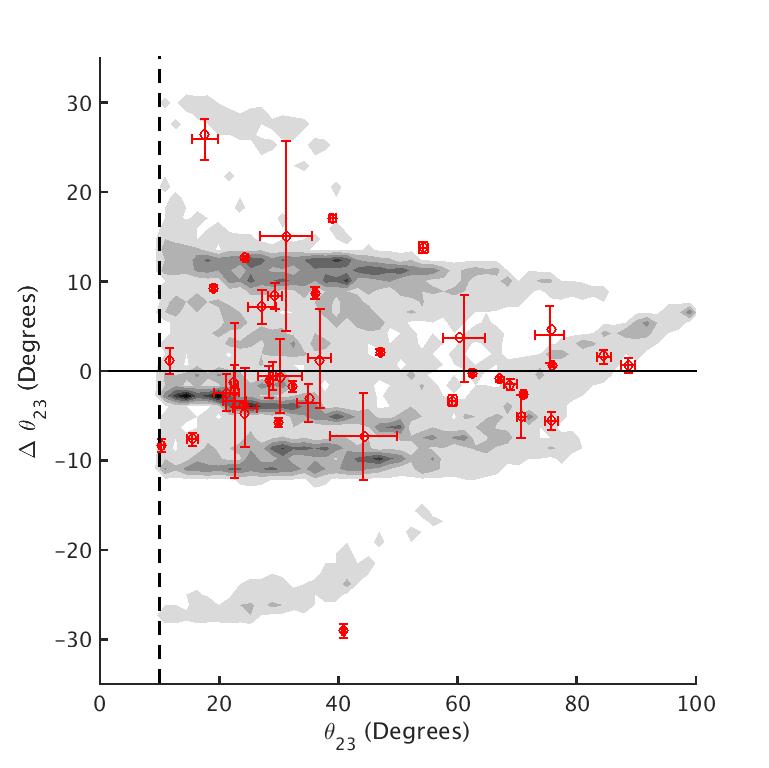}  
  \label{fig:sfigtmagbias}
\end{subfigure}
\caption{Deviations from the FSQ, projected along the \(\theta_{23}\) axis, are plotted for a single lens shown in the left panels of Figure \ref{fig:10xCDMcontour}.
        Here, the left panel shows a scatter plot of quads, while the middle and right panels show simulated quads by density.
        Grayscale corresponds to the simulated quad population while red circles with error bars corresponds to observations.
        This shaded plot style will be used from now on, since it is easier to display densities of points, particularly when the number of 
        points is large. Quads with \(\theta_{23}<10^{\circ}\) are omitted with the cutoff shown as a vertical dashed line.
        Interestingly, for this particular lens, there exist some quads with \(\theta_{23}>90^{\circ}\), which normally 
        does not happen. This can occur if one of the images lies near a particularly large subhalo, which delays that image and changes the arrival time order.
        This only happens for a handful of quads in only the most extreme lenses and only for \(10\times\Lambda\)CDM. It is not a concern in our analysis. 
        The population of quads in the right panel is the same population, after the 
        most effective selection bias from Table \ref{table:tmag} is applied (See Section \ref{sec:exbias}). As before, the vertical dashed line represents the cutoff removing quads with 
        \(\theta_{23}<10^{\circ}\). This particular bias uses the 0th percentile (minimum of the data) as \(\xi_1\) and the 75th percentile as \(\xi_2\). The biased population is 
        consistent with the observed population (p=32\%) while the unbiased is not (p=4.1\%). This is mostly because the bias has made 
        quads with lower \(\theta_{23}\) more likely, moving the denser part of the gray population to the left making 
        it more consistent with the denser part of the red population.}
\label{fig:sing10xcdmfsq}
\end{figure*}

The deviations from the FSQ for the simulated lens are plotted in Figure \ref{fig:sing10xcdmfsq}. The deviations are much larger now
and are of the same order as observations, so now more care is necessary to confirm or rule out consistency.
The metric we use to test 
consistency is the two-dimensional Kolmogorov-Smirnoff (KS) test originally presented by \citet{peacock83}, expanded on by \citet{fasano87}, 
and made readily available by \citet{numrec}.
Unlike the one-dimensional KS test, the 2D KS test is not strictly independent of the shape of the distribution 
because the Cumulative Distribution Function (CDF) is not uniquely defined in more than one dimension \citep{peacock83}. Fortunately, this effect is minuscule for cases 
where the x and y values for the data set are not strongly correlated, and is not measurable in the case of our data.

The test works by taking a point from one of the distributions 
and counting the fractional number of points from each distribution in each of the quadrants around the starting point. It then repeats this 
process using each point as its starting point and finding the one which creates the quadrant with the maximum discrepancy between the two
distributions. This discrepancy can then be evaluated for statistical significance \citep{peacock83,fasano87}.
Like the traditional KS test, the 2D KS test returns a p-value which is the probability
of obtaining a more extreme discrepancy than measured, assuming the null hypothesis is true. In this case, the null hypothesis is the claim that 
the observed population is simply drawn randomly from the synthesized population.
For our purposes, a p-value less than 5\% will 
indicate that the simulated population and observed population are indeed different. 

The uncertainites in observations are accounted for by taking each observed quad
and replacing it with 100 points distributed in a 2D Gaussian with \(\sigma_x\) and \(\sigma_y\) corresponding to the astrometric uncertainties in observations and
centered on the intersection of the error bars. This spreads out the density of points in accordance with the error bars.
Since this process artificially gives the population 100 times the number of points, for the purpose of testing statistical significance 
it is still considered to consist of only the 39 quads.

When applied to the simulated quads from the lens in Figure \ref{fig:sing10xcdmfsq},
the 2D KS test returns a p-value of 4.1\%, meaning that the population of quads produced by this single synthetic lens is almost consistent with the observed population of quads.

\begin{figure}
\includegraphics[width=\linewidth]{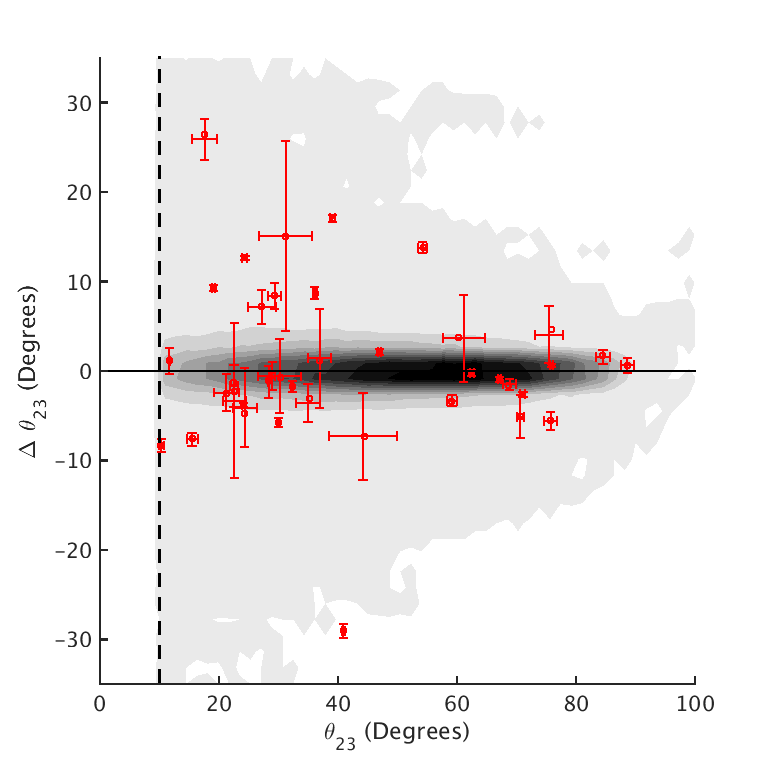}
\centering
\caption{Deviations from the FSQ for the population of over 100,000 quads from 200 synthetic galaxies with \(10\times\Lambda\)CDM substructure and no external shear (grayscale),
         compared to observations (red).
         Since the majority of the 200 galaxies have few large perturbations from elliptical contours, only small deviation from the FSQ results when
         considering a population.} 
\label{fig:10xCDMpop}
\end{figure}

We should not necessarily expect that any single lens would be able to reproduce the entire population of observed quads, since they themselves come
from many different galaxies. Instead it makes more sense to compare a population of simulated lenses
with the observed quad population. Specifically we created 200 lenses 
with different random seeds in the manner described in the previous section and plotted their deviations from the FSQ in Figure \ref{fig:10xCDMpop}. It is immediately apparent
that the deviations from the FSQ are even less than that for the single lens. This is because there are more lenses similar to the lens on the right in Figure \ref{fig:10xCDMcontour}
than that on the left in the same figure, which has larger perturbations.
This makes the number of quads that deviate from the FSQ more diluted and the population an even weaker match to observations. 
The 2D KS test confirms this, returning a p-value of \(0.099\%\).
This indicates that even with \(10\times\Lambda\)CDM substructure, the observed deviations 
from the FSQ cannot be explained. 

We consider this experiment to be the most critical of the experiments done
in this paper, and therefore have committed the computational resources necessary to synthesize 200 lenses. Other populations within this paper are synthesized using less than 200 lenses. 
Curious about whether a population of 20 galaxies would return the same p-value as one of 200, we divided the 200 lenses into 10 sets of 20 and calculated the p-value 10 times. 
Values returned typically ranged from 0.03\% to 0.23\%. We interpret this to mean that when only 20 galaxies are used in other tests, the p-value can vary by $\sim0.1\%$ (for the unbiased case- 
when selection biases are applied in Section \ref{sec:biases} this value will be different).

\subsection{Nonzero external shear}\label{ssec:shear}
External shear is a common feature in lens models because it is easy to express analytically and it seems to fit many lenses, 
although it may not necessarily correspond to a readily identifiable physical entity. 
\citet{wong2011} found disagreement between fitted values for external
shear in lens models and measured values from the environments of those lenses, indicating that the external shear inferred from model lenses 
in reality corresponds to a handful of environmental factors including not only external shear, but also line-of-sight effects,
as well as compensates for simplifying assumptions about the main lens.
Rather than thinking of external shear as a physical observable quantity, perhaps it makes most 
sense to instead think of it as a simple first order fitting parameter that represents 
information from several unknown effects. Whatever the case, \citet{FSQ2} document the effects of external shear on the distribution 
of quads relative to the FSQ. 
Since this shear can provide deviations from the FSQ, we can experiment with 
nonzero values of shear and see if this is able to better match observations.

\citet{slacs5} modeled 63 lenses discovered in the Sloan Lens ACS Survey (SLACS) and fit values for the external shears of each lens, 
ranging from 0 to 0.27 with a median of 0.05. 
Since the authors argue that their population of lenses is statistically consistent with being drawn 
at random from the survey, we can assume that the shears and axis ratios they found are representative of typical
external shear values for these types of lenses. The distribution of shear values from SLACS is consistent with the values determined from lens environments \citep{wong2011},
and the distribution of axis ratios from SLACS is consistent with that of nearby ellipticals \citep{ryden92}. Both of these consistency checks come from methods which are independent
from lensing models. It is therefore justified to use these values when synthesizing a population of lenses. For the
same 63 lenses, the authors also list the axis ratios for each lens from their model, ranging from 0.51 to 0.97 with a median of 0.79, providing a natural
way to make the synthesized population of main halos more representative of a true population.

We created a population of 20 lenses. The subhalos are made the same way as above, with \(10\times\Lambda\)CDM substructure, while the main halos have axis ratios and external 
shears randomly drawn from the 63 values in \citet{slacs5} and given a random shear orientation angle. Since the axis ratios are not all the same, the caustic size
differs on a lens-by-lens basis. This is because in the limiting case of axis ratio = 1 the inner caustic becomes a point, so nearly circular lenses have smaller diamond caustics 
than elliptical ones. This means the lensing cross section for quads is different for each galaxy, so it no longer makes sense to construct a population
with simply 10,000 quads for each lens. Instead, the number of quads for each lens is proportional to the caustic area. 
Once again the deviations from the FSQ are too small to match observations.  The 2D KS test
confirms this, returning a p-value of \(0.0077\%\). Even the combination of \(10\times\Lambda\)CDM substructure, realistic external shears, and realistic axis ratios are unable
to produce the deviations from the FSQ necessary to explain observations.

\section{Introducing quad selection biases}\label{sec:biases} 
Observational selection biases affect all surveys. The quad sample we are using in this paper is very heterogeneous: some quads were discovered as part of a well-defined survey 
while others were discovered individually. This means that correctly accounting for biases is impossible. In lieu of a known selection function, we have devised a makeshift model
which biases quad selection in a probabalistic sense. In the future, quads will be discovered by the Large Synoptic Survey Telescope (LSST), with well-defined selection criteria. 
In the meantime, our treatment is sufficient to mimic selection effects and gain intuition as to their general implications.

In Section \ref{sec:ellpert} populations of quads will be generated which are closer to the observed distribution and so selection biases will be important. 
As such, we consider these biases now, and apply them to the population of quads generated in Sections \ref{ssec:10xCDM} and \ref{ssec:shear}.

Three main biases are considered.
First, quads which are brighter are more likely to be detected. The source luminosity is 
uncorrelated with the lens properties and is therefore not relevant to this analysis. What matters in this context is
the total magnification of all the images. If the images are highly magnified, the quad will likely be detected. Another potential source for bias 
is the separation between images \citep{oguri2006,pindor2003}. If 2, 3, or even all 4 images are close together, they may not be resolved as distinct images and the quad may instead 
look like a triple, double or a point source.
Such a case would not be included in the observed galaxy quad population. Finally, quads which have a large contrast between
the magnification of the brightest image and the dimmest will also be less likely to be resolved as having distinct images \citep{oguri2006,pindor2003}. 
These cases are unlikely to register as more than
a point source in a survey and may not be followed up with deeper observations. These inherent biases in the way quads are observed could 
select quads whose properties are different from those of an unbiased population. It is not hard to imagine this affecting the distribution of quads around 
the FSQ, so we examined the consequences of these biases.

Optimally one would simply know the limiting resolution and magnitude of one's survey and omit synthetic quads that are outside of that range. However, 
the population of known quads comes from an amalgam of many different surveys, making it difficult to systematically identify the degree of lensing 
biases \citep{oguri2006}.
Although quads discovered may have a biased population due to inability to resolve closely spaced images or images of drastically different magnifications, 
this paper will largely ignore these biases in favor of the total magnification bias. The main reason for this is that the central question 
of this exercise asks if it is at all possible that biases could explain the observed population of quads via selection effects. 
When addressing this question, it makes sense to look at the populations in the best possible light, and our tests seem to 
indicate that the total magnification bias results in 
the highest p-values compared to the flux ratio and image separation biases. From now on, discussion of biases will be limited to the total magnification bias.

\begin{figure}
\includegraphics[width=\linewidth]{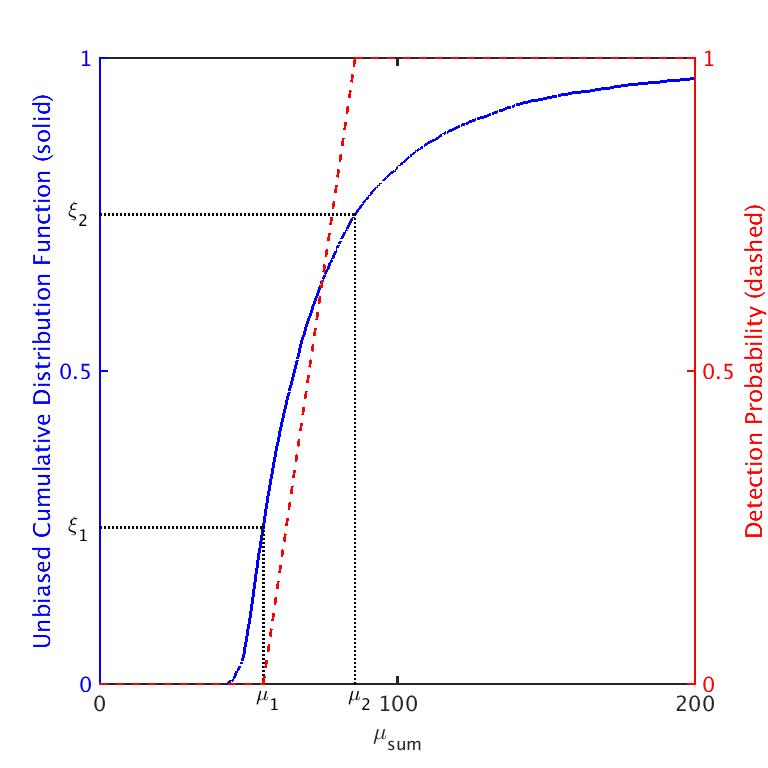}
\centering
\caption{A schematic presentation of how the bias is applied. The blue solid curve shows the CDF of the magnification (summed from all four images) for the unbiased quad population.
         The red dashed line represents the probability that a quad will be detected and kept in the population after the bias is applied. 
         The minimum threshold value below which detection is impossible, \(\mu_1\), is \(\xi_1\), the 25th percentile in this example. 
         Likewise the maximum threshold value above which detection is certain, \(\mu_2\), is \(\xi_2\), the 75th percentile here.
         Once the thresholds are set, the probability of detection scales linearly with the value of the summed magnification between the thresholds, shown by
         the red dashed line.} 
\label{fig:cdfex}
\end{figure}

To properly account for the magnification bias, one would need knowledge of the quasar luminosity function and lens mass distributions for all redshifts \citep{han15}.
Even then, it is infeasible to identify any particular magnitude cutoff that applies to all surveys. 
Conscious of our ignorance, we use the following logic to approximate the general effects of the bias. 
First, imagine that above (below) a certain threshold for the summed image magnifications a quad is guaranteed to be (not be) detected.
Between these two thresholds, suppose the probability of detection scales linearly with the summed magnification.
Since those thresholds are difficult to pinpoint, we set them to a percentile of the data, \(\xi_1\) and \(\xi_2\), e.g. the 25th percentile and 75th percentile. 
This process is shown visually in Figure \ref{fig:cdfex}. The exact value
of these percentiles is unknown in actual surveys and the effects of changing them will be an important part of analyzing the effects of the bias. 
The application of this bias occurs before the \(\theta_{23}<10^{\circ}\) cutoff selection. Our bias will be applied to all populations presented in following sections.
\subsection{Example Bias}\label{sec:exbias} 
Before applying the bias to all synthesized quad populations, we will first apply it to a single population to convey its general effects. The population 
that we believe depicts this best is that of the first single lens with \(10\times\Lambda\)CDM subhalos, (Figure \ref{fig:sing10xcdmfsq}). 
This is because this population has appreciable deviations from the FSQ leading to visible distributions 
in both the \(\theta_{23}\) and \(\Delta\theta_{23}\) dimensions. 
It will turn out that the bias notably affects only the 
distribution in the \(\theta_{23}\) dimension, but this is most readily seen when the spread of points in both dimensions is large. 
The trends we see here apply to all populations of quads.

The bias makes the quads with the larger total magnification
more likely to be detected, while throwing out the fainter quads which are unlikely to be detected. Table \ref{table:tmag} shows the resulting p-values from the 2D KS test after this
bias has been applied with various values for minimum percentile, \(\xi_1\), below which detection is impossible and the maximum percentile, \(\xi_2\), above which detection is certain.  
Since the strength of the bias is unknown, we will imagine it is as effective 
as it possibly could be in making the observed and synthetic populations consistent with one another. 
The highest p-value corresponds to the best-case scenario. With this being said, we did not feel it was necessary to run an exhaustive search for the maximum because 
the exact specificity in threshold percentiles chosen is not particularly useful. Instead, we simply use the highest p-value in Table \ref{table:tmag},
\footnote{The 99th percentile is chosen as the highest \(\xi_2\) rather than the 100th percentile 
  because if the latter is used, a single large-magnification outlier, which would likely be code-resolution artifact, could 
  drastically decrease the slope of the linear detectability function (Figure \ref{fig:cdfex}) and artificially cause nearly all quads to have a low probability of detection.}
which is 32\%. This value is greater than the 5\% significance threshold, meaning that, when constructed from this single lens with an optimistic 
bias applied, this particular population of quads is consistent with observations.
\begin{table}
\centering
\setlength\tabcolsep{4pt}
\begin{tabular}{c c |c | c c c c c }
\multicolumn{2}{c}{}& &\multicolumn{5}{c}{Single Lens Example:}\\ 
\multicolumn{2}{c}{}& &\multicolumn{5}{c}{p-values for Biases (\%)}\\  
{\multirow{5}{*}{\(\xi_1\)}} & 99 &\vline& -- & -- & -- & -- & --  \\ 
			     & 75 &\vline& -- & -- & -- & -- & \(0.0002\)  \\ 
			     & 50 &\vline& -- & -- & -- & \(2.8\)& \(0.033\)  \\ 
			     & 25 &\vline& -- & -- & \(29\)& \(13\)& \(0.46\)  \\
			     &  0 &\vline& -- & \(8.8\)& \(26\)& \(32\)& \(4.2\)  \\ \cline{1-2}\cline{3-8}
 \multicolumn{2}{c|}{}&\vline& 0 & 25 & 50 & 75 & 99  \\ 
 \multicolumn{2}{c|}{}&\vline& \multicolumn{5}{c}{\(\xi_2\)}  \\ 
\end{tabular}
\caption{Optimizing the effect of the selection bias. P-value results are presented for various minimum and maximum cutoffs for the total magnification bias. p-values greater 
  than 5\% indicate that the populations are consistent with the null hypothesis, which claims the observed population 
  comes from selecting quads from the
  synthesized population. For this example lens there are several cases with such p-values, with the highest value ocurring in the case where the minimum cutoff for the 
  bias is the minimum of the data and the threshold for certain detection is the 75th percentile of the data.
  This case puts the population in the most positive light, with 5177 out of 8907 quads detected, and is plotted in Figure \ref{fig:sing10xcdmfsq} (right panel).
  The number of quads remaining in the population after the bias is applied is not 
  depicted, but is lowest in the upper right corner (312 remain out of 8907) and highest in the bottom left. (8233 remain) }
\label{table:tmag}
\end{table}

\begin{figure}
  \includegraphics[width=\linewidth]{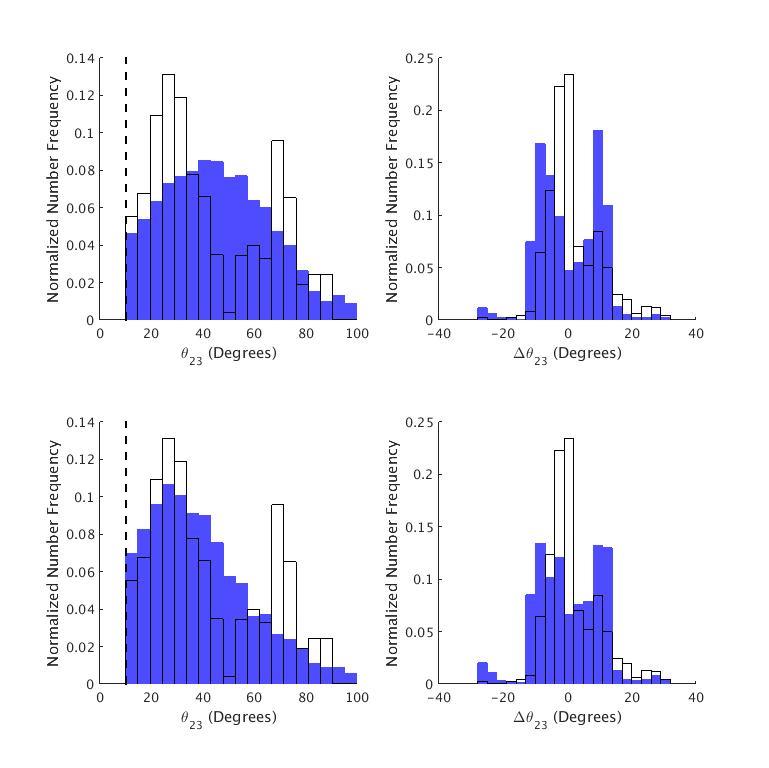}
  \centering  
\caption{Marginalized distributions are shown for both the \(\theta_{23}\) (x-axis of Figure \ref{fig:sing10xcdmfsq}) and \(\Delta\theta_{23}\) (y-axis of Figure \ref{fig:sing10xcdmfsq})
         for the population of quads
         created from the single \(10\times\Lambda\)CDM lens (blue), compared to the observed population (open white histogram).
         The top two panels show the unbiased population while the panels on the bottom show the biased population which
         yields the best match to observations. The cutoff removing quads with \(\theta_{23}<10^{\circ}\) is shown as the dashed line.
         The bias strongly skews the \(\theta_{23}\) distribution to lower values, matching
         up more closely with the larger peak in the observed data, which is the largest factor in why 
         the p-value improves. Meanwhile the general shape of the \(\Delta\theta_{23}\) distribution is only slightly affected 
         by the bias.}
  \label{fig:tmagbiasmarg}
\end{figure}

The effect of this bias is shown in Figure \ref{fig:sing10xcdmfsq}, but is perhaps easier to see in the marginalized distributions, shown in Figure \ref{fig:tmagbiasmarg}. 
The bias has little effect on the distribution in \(\Delta\theta_{23}\) but a significant effect on the distribution in \(\theta_{23}\). Both figures show that 
the bias selectively removes quads in a way that shifts the remaining population to smaller \(\theta_{23}\).
This makes sense because smaller \(\theta_{23}\) means that the 2nd and 3rd arriving images are close together, which comes from the quads where the source 
is near the caustic line, resulting in high magnification. These quads are more magnified and more likely to remain after the bias is applied.

Though we have only depicted the case returning the highest p-value, in reality there are many different realizations possible in which the distributions, biased at different levels, 
would take on intermediate forms between the two cases depicted in the middle and rightmost panels of Figure \ref{fig:sing10xcdmfsq}. 
It is also possible to have a stronger bias with more pronounced
effects than those shown in the right panel, but this would result in a lower p-value than what is shown here.

Independent of the strength of the bias, it is important to note that the bias strongly affects the \(\theta_{23}\) 
distribution and only weakly affects the \(\Delta\theta_{23}\) distribution.
The result is that the \(\Delta\theta_{23}\) distribution is the more important one when attempting to decipher whether or not a population is consistent with observations: 
if there is a mismatch between the synthesized population's \(\theta_{23}\) distribution and that of the observations, there may exist a bias that could bring the population into the
realm of plausible consistency, however 
if there is a similar mismatch between the synthesized population's \(\Delta\theta_{23}\) distribution and that of the observations, no bias will fix the problem.
\subsection{Population Bias Results} 
Now that the effects of such a selection bias have been demonstrated for a single lens, it is time to apply the same bias to the population of lenses which is meant to represent 
the galaxy lens population. The table analogous to Table \ref{table:tmag} is not presented, but the same analysis is run, this time for the population of 200 lenses with 
\(10\times\Lambda\)CDM substructure with no shear. The most optimistic bias leaves the simulated quad population with a p-value of \(1.3\%\), which is still inconsistent with the null hypothesis. 
A similar exercise as in Section \ref{ssec:10xCDM} --where we recalculate p-values with 10 subsamples using 20 galaxies each-- yields p-values that typically vary from 0.8\% to 2.3\%.
This means for populations using only 20 lenses, the p-value can be expected to vary by $\sim1\%$ in the biased case.
Using the population of 20 lenses from the case with nonzero shear (Section \ref{ssec:shear}), the best match occurs under the same cutoff values, again with a p-value of 1.3\%. 
Even when the most optimistic bias is applied, 
these synthesized populations from the \(10\times\Lambda\)CDM substructure scenario are completely inconsistent with observations. It therefore seems 
unlikely for \(\Lambda\)CDM substructure to account for the observed deviations from the FSQ.

\section{Deviations from elliptical lenses}\label{sec:ellpert}
Aside from \(\Lambda\)CDM substructure, there are other effects which are capable of producing asymmetry in the lens to create significant deviations from the FSQ. 
If the mass of the galaxy is not relaxed into a single smooth profile, for example, then there could be inherent asymmetries. \citet{chae14} found that in order to fit the 
density profiles for early-type galaxies a two-component mass model was required. \citet{Young16} found that simulations of both dark matter only and dark matter + hydrodynamics 
resulted in mass distributions which are not fully relaxed.
If the baryons and dark matter do not coalesce into identical distributions then it would be possible to have a galaxy which has two related
but not identical distributions. With two non-identical distributions, there are several possible realizations which could break the double-mirror symmetry or give rise
to ``wavy'' features in the lens projected isodensity contours. 

\begin{figure}
  \includegraphics[width=\linewidth]{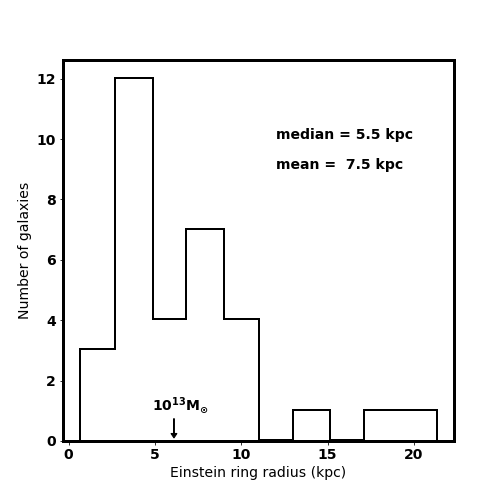}
  \centering  
  \caption{Comparison of the baryon-dark matter transition radius with the Einstein radius. The histogram displays the Einstein radius for the 33 observed quads which have known lens redshift.
           The average and median of this distribution are presented in the figure.
           The radius of where the dark matter component becomes dominant over the baryon component depicted here is calculated 
           in \citet{chae14} for a \(10^{13}\text{M}_{\odot}\) halo, a typical mass for galaxies which host quads.            .
           It is fortuitous that these radii values should coincide.}
  \label{fig:transrad}
\end{figure}

The quad image circle in halos of \(\simeq10^{13}\text{M}_{\odot}\), slightly larger than that in our halos, has a radius of around 6 kpc. 
It is a remarkable coincidence that this radius happens to correspond to a transition region
from baryons to dark matter, illustrated in Figure \ref{fig:transrad}. At smaller radii, the baryons are the dominant mass component while the dark matter dominates at outer radii.
It just so happens that the radius where they have comparable mass lies at a similar radius as the position of images, 
which is fortunate because the image circle is the region lensing can most precisely probe. 
It also complicates matters, because if there are any inherent asymmetries in the baryon and dark matter distributions, this transition area
is where they will have the most drastic effects on image positions.
It stands to reason that various perturbations arising from ellipticity transitioning between the baryon and dark matter distributions could result in deviations from the FSQ.
This motivates an additional series of experiments. 

This time, we construct lenses with no \(\Lambda\)CDM substructure, but using two superimposed elliptical Einasto profiles instead of just one. 
The first profile represents the dark matter, in which shape parameter \(\alpha\) is changed to 0.18 but otherwise the same as before. This slightly larger shape parameter
makes the profile less concentrated than before, making the scale radius 13.6 kpc. The central density for the profile representing the baryons is set to 
\(5\times\) the dark matter \(\Sigma_0\), motivated by the Illustris simulations \citep{Vogel14,Young17}, 
but the profile drops off much more steeply than that of the dark matter. 
The baryon profile is given a scale radius of 1 kpc and a shape parameter of 0.6, which have been chosen to make the slope near the image radius more realistic, as in Section \ref{ssec:math}. 
This means the baryons will be the dominant mass component in the very central regions, but at radii near the images the dark matter has become dominant.

\begin{figure}
  \includegraphics[width=\linewidth]{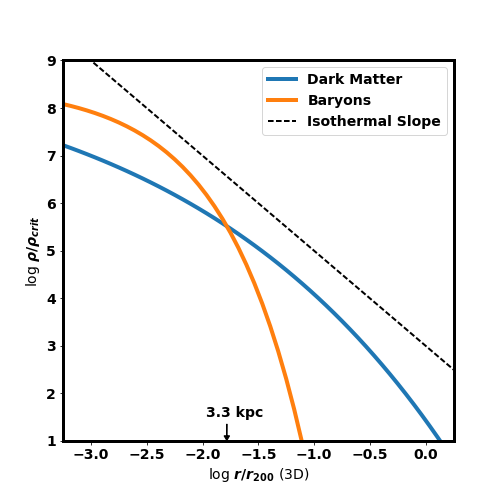}
  \centering  
  \caption{Density profile for the two components from which lenses will be constructed. The blue solid line represents dark matter and orange represents baryons. 
           The dashed line depicts an isothermal slope. The arrow indicates the transition radius where the densities are equal, at 3.3 kpc. 
           The axes are scaled with respect to the critical density of the universe at lens redshift 
           (z=0.6) and \(r_{200}\) so that the graph can be readily compared with Figure 12 of \citet{chae14}. It should be noted that this is the analytical
           form from Equation \ref{eq:einastoprof} and is spherically symmetric. Once ellipticities and perturbations from ellipticities are applied (Section \ref{sec:ellpert}), 
           the true profile will differ as a function of position angle, but should be reasonably close to this when spherically averaged.}
  \label{fig:twoprofile}
\end{figure}

The (3D) transition radius in this setup actually lies near 3 kpc instead of 6 kpc, which is mostly due to the smaller halo size. The transition radius for a 
\(10^{12.5}\text{M}_{\odot}\) galaxy, approximately the mass of our halos, is closer to 4 kpc \citep{chae14}. We consider this slight mismatch between 
3 and 4 kpc acceptable, recognizing that if the radii matched better, any asymmetries 
created would have a more pronounced effect on quad image angles. This transition radius is identifiable in Figure \ref{fig:twoprofile}, which shows 
the density as a function of radius for the two components.

Within the simulation window, \(\simeq25\%\) of the mass is in baryons. At infinity, only about 4\% of the mass is in baryons. Depending slightly on the exact values for the ellipticity, 
the Einstein radius increases to \(\simeq5.2\) kpc due to this additional baryonic mass. Axis ratios for the elliptical profiles are separately drawn from \citet{slacs5}.
As before, the number of quads per galaxy is again proportional to the caustic size, and quads with \(\theta_{23}<10^{\circ}\) will again be omitted. External shear is left at zero.
Perturbations will be applied to the elliptical structure of the two profiles-- dark matter and baryons-- as described in the next sections.

\subsection{Fourier Component Perturbations} \label{ssec:fourier}
The first form of mass perturbations from pure ellipses we examined are motivated by observations. \citet{bender87} 
measured deviations from ellipticity in isophotes in terms of Fourier expansion in the polar angle. 
Radial isophote deviations from a perfect ellipse are parameterized by coefficients of sines and cosines,
\begin{equation*}
    \Delta R=\sum_{k=3}^{6} a_k\cos(k\phi)+b_k\sin(k\phi) ,
\end{equation*}
where \(\phi\) is the angle with respect to the ellipse axis. The index starts at 3 because \(a_1\), \(a_2\), \(b_1\), and 
\(b_2\) are already constrained by ellipse parameters such as 
axis ratios, semimajor axis, and center position. Since then, numerous studies have
followed their notation. Though in principle there could exist higher order deviations, the 
index is typically cut off at \(k=6.\) The values of the coefficients change for each isophote and are therefore 
a function of radius, but this dependence is complicated and for our purposes we will just use average values.
The most commonly discussed deviation is the \(a_4\) term. Positive values deform the 
ellipse into a diamond-shaped ``disky'' isophote while negative values deform the ellipse into a 
``boxy'' shape. \citet{mitsuda16} analyzed the \(a_4\) values for early-type galaxies at \(z\sim1\) and 0, finding 
values roughly distributed from -0.02 to +0.04.
\citet{corsini16} measured the Fourier coefficients for three nearby galaxies and found \(a_4\) to range from -0.03 
to 0 in one case, between 0 and 0.06 in another case, and 0 and 0.03 in the third case.
They found \(a_3\) and \(b_4\) to be largely consistent with 0,  but in one case had \(b_3\) and \(b_6\) range from 0 
to about 0.03 and two cases where \(a_6\) ranges from 0 to approximately 0.03.
They did not include information as to \(a_5\) or \(b_5\). \citet{kormendy09} found values of \(a_4\) typically between 
-0.02 to 0.02 for galaxies in the Virgo cluster, where \(a_4\) was as large as 0.09 in one case. 
Nonzero \(a_3\) values were also found, but were not as extreme as the \(a_4\) values measured.
These general results give an impression for the order of deviation from ellipticity for realistic galaxies and provide 
a framework from which to construct a galaxy population.

We construct this population by having \(a_4\) uniformly selected between -0.04 and 0.04 and \(a_6\) uniformly selected between \(-0.02\) and \(0.02\). Unlike real galaxies, the 
selected value for the Fourier coefficients is kept constant as a function of radius, but should still provide insight into the effects these deviations from ellipticity have on quad
deviations from the FSQ. Though the observations above only apply to the light, it is assumed in this context that the dark matter profiles likewise have deviations from ellipticity of 
similar order. As such, both profiles get different values chosen for \(a_4\) and \(a_6\). The other coefficients are left at zero. For now, the major axes of the two elliptical
distributions are colinear and the centers coincide.

One hundred galaxy lenses are constructed and the resulting population of quads is analyzed using the 2D KS test. 
\footnote{One hundred lenses are now used to create a population as opposed to twenty used for most of the cases before because as more types of perturbations 
   are added, parameter space gets larger than before, so having more lenses in a population is necessary.}
The result is a p-value of 0.00044\% in the unbiased case and 0.013\% when the same bias as before is applied. This is less than the 5\% threshold,
indicating that the Fourier perturbations alone are insufficient to match observations.

\subsection{Misaligned Ellipses}\label{ssec:misaligned}
Another case to explore is that in which the baryons and the dark matter have elliptical projected mass distributions but their ellipse major axes do not 
necessarily line up perfectly i.e. the two distributions have different position angles (PAs). 
Perhaps a population of lenses with features like this could be responsible for the observed deviations from the FSQ. In this test,
the dark matter profile has an ellipse PA that is tilted with respect to the x-axis by an angle 
randomly selected between 0 and 45 degrees. The PA of the baryon profile remains aligned with the x-axis. Fourier coefficients introduced in Section \ref{ssec:fourier} are set to zero.
A population of quads is synthesized from 100 lenses like this and subjected to the 2D KS test.
Before biasing, the p-value returned is 0.0017\% and after the bias is applied the p-value returned is 0.020\%.
Thus, the effect of misaligned ellipses is also nowhere near sufficient to create the necessary deviations from the FSQ.

Another variant of this idea of having two elliptical profiles give rise to a non-elliptical mass distribution is to offset the centers of the ellipses themselves. 
Image positions are most sensitive to mass perturbations near the image radius, 
so it is the perturbations near the image radii that we truly wish to emulate.
If the centers of the profiles were not coincident it would cause a non-elliptical perturbation even at radii farther out than the centers. 
In this way, artificially offsetting the centers can serve as an easy-to-generate asymmetry.
In reality, the centers of baryonic and dark matter distributions are thought to be nearly coincident. 
In lensing models it is usually assumed that the centers coincide \citep{slacs4,slacs5}. However, since the positions of quads are more sensitive to structure 
at the image radius than structure in the lens center, we will accept that this model is inaccurate in most central region of the lens in return for the structure
at image radius it generates.
That is to say, we will offset the centers of the two distributions in the simulations, but this is not necessarily a claim that the centers are in 
reality offset so drastically. 
Instead, the offset centers are an artificial way to introduce non-ellipticity near the image radius, which could exist in real lenses.
Having said that, we note that it is possible for the centers of dark matter and baryonic distributions to be non-coincident, within the framework of 
self-interacting dark matter \citep{Kahl14,Kahl15}.
Offsetting these centers could be thought of as additional Fourier perturbations of lower order than 3, since these coefficients are constrained 
by the ellipse center position among other parameters.

We create a population consisting of 100 simulated galaxies, each with a baryon and dark matter component described above. The ellipse PAs of the two distributions are again tilted by an 
angle between 0 and 45 degrees while the centers are offset by a radius randomly selected between 0 and 15 pixels (1 kpc) in a random direction in the 2D lens plane. 
The center of the lens is considered to be the center of the baryon distribution since this is the distribution which an observer would see and assume to be the center.
The 2D KS test for the population compared to observations results in a p-value  of 0.053\%. After the bias, the KS test returns a p-value of 3.4\%, still indicating inconsistency with 
the null hypothesis. 

\subsection{Combined Effects}\label{ssec:combo}
\begin{figure*}
  \centering
  \begin{tabular}[c]{cc}
    \begin{subfigure}[c]{0.34\textwidth}
      \includegraphics[width=\linewidth]{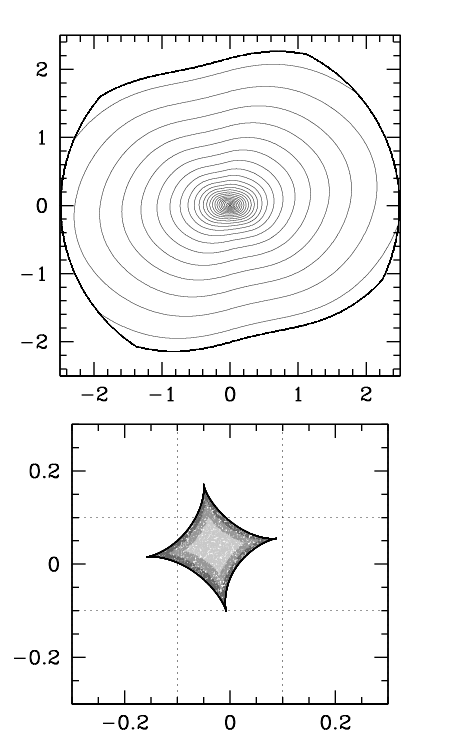}
      \label{fig:sfigcontour1}
    \end{subfigure}&
    \begin{subfigure}[c]{0.34\textwidth}
      \includegraphics[width=\linewidth]{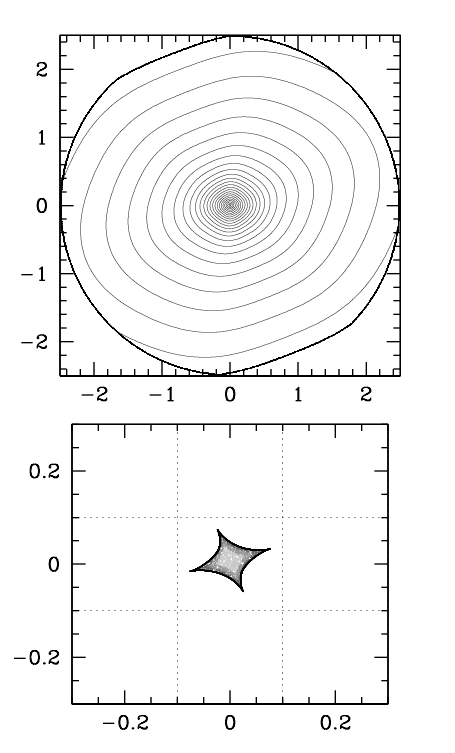}
      \label{fig:sfigcontour2}
    \end{subfigure}\\
    \begin{subfigure}[c]{0.34\textwidth}
      \includegraphics[width=\linewidth]{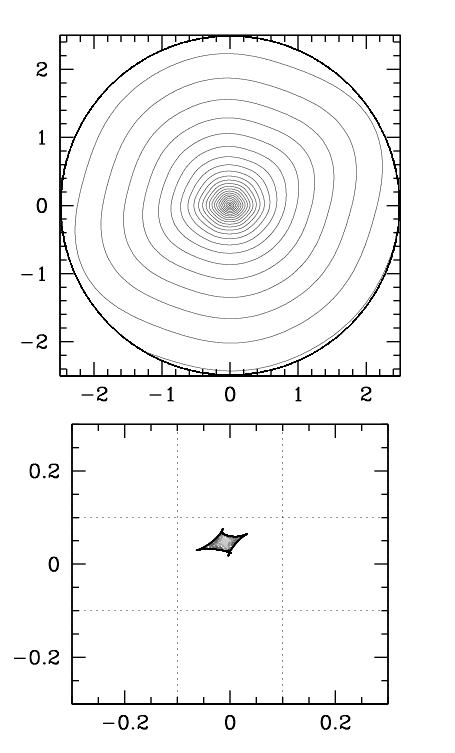}
      \label{fig:sfigcontour3}
    \end{subfigure}&
    \begin{subfigure}[c]{0.34\textwidth}
      \includegraphics[width=\linewidth]{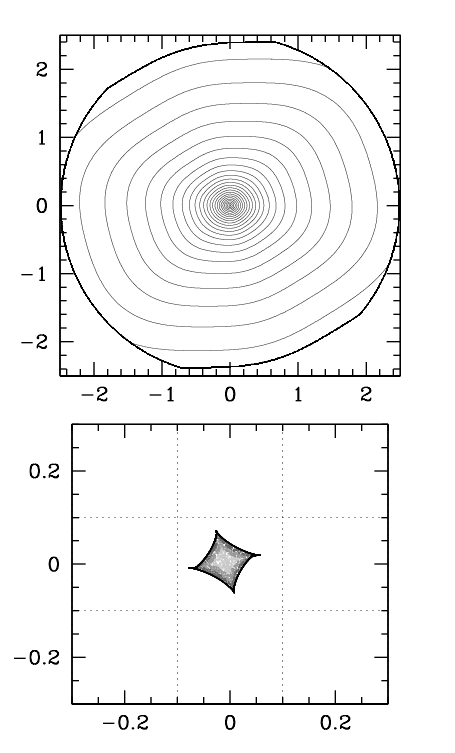}
      \label{fig:sfigcontour4}
    \end{subfigure}\\
  \end{tabular}    
\caption{Four example lenses and caustics from the population which includes offset centers, 
         tilted ellipse axes, and nonzero \(a_4\) and \(a_6\) for both the dark matter and baryon distributions.
         Significant deviations from ellipticity are clear, which produce noticable changes in the caustic 
         and thereby the image positions and deviations from the FSQ. Note the resemblance of these mass density contours to the shape of
         the observed isophotes of \citet{mitsuda16}. The thick curve is the boundary outside of which the mass is set to zero. 
         Its shape arises from the combination of a cut along a particular isodensity contour and a circular cut at the edge of the 
         simulation window (16.7kpc). 1 arcsecond corresponds to 6.7 kpc. }      
\label{fig:contourpopex}
\end{figure*}

Each of the types of perturbations described in Sections \ref{ssec:fourier} \& \ref{ssec:misaligned} carries potential FSQ deviations with them. It would be remiss not to test the combination of effects.
This time 1000 lenses are created with axis ratios from \citet{slacs5} for both the dark matter and baryon distributions. The elliptical distributions are tilted with respect to one another 
by an angle between 0 and 45 degrees. The center for the dark matter distribution is offset by between 0 and 1 kpc, and each distribution shape is altered via \(a_4\) (between -0.04 and 0.04) 
and \(a_6\) (between -0.02 and 0.02). Four example galaxies are shown in Figure \ref{fig:contourpopex}. We invite the reader to compare these synthetic distributions
to observed galaxies in Figure 12 of \citet{mitsuda16}, which bear visual resemblance to one another. 

\begin{figure}
  \centering
  \includegraphics[width=\linewidth]{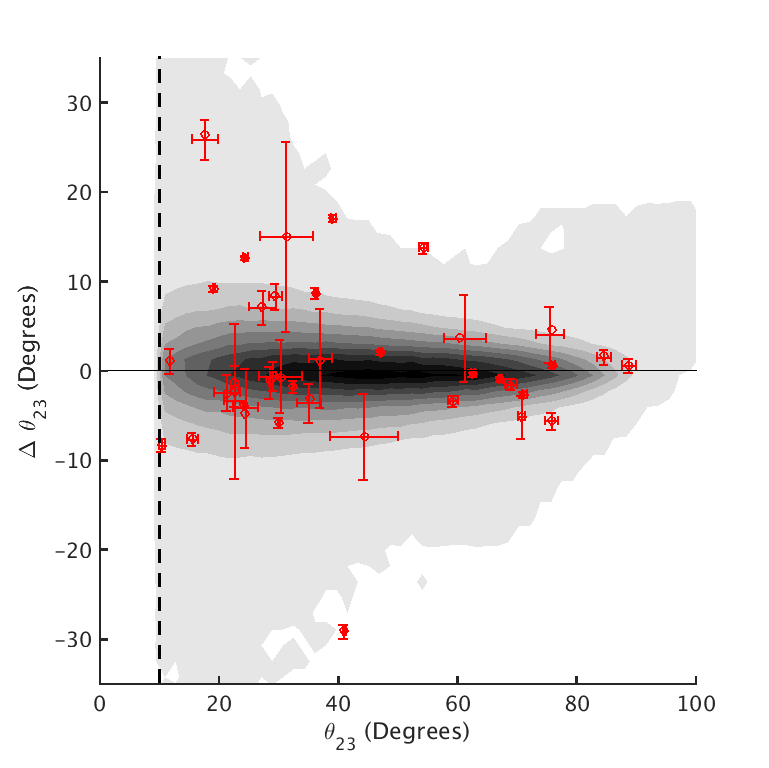}  
\caption{Deviations from the FSQ generated from the population of galaxies like those in Figure \ref{fig:contourpopex}, with the most optimistic
         bias applied. The deviations from the FSQ are consistent with observations with a p-value of 6.2\%.
         } 
\label{fig:ellpert}
\end{figure}
The deviations from the FSQ generated by this population of galaxies are shown in Figure \ref{fig:ellpert}. The 2D KS test results in a p-value of 0.16\% for the unbiased 
distribution and 6.2\% once the bias is applied.
This result implies that a combination of the above perturbations from pure ellipticity could be consistent with observations if observational biases are favorable.

\section{Discussion and Conclusions}\label{sec:conclusion}
We have sought to construct a population of quadruple image systems, generated by synthetic  lensing galaxies, which is consistent with the observed distribution of quads 
relative to the Fundamental Surface of Quads (FSQ) (Figure \ref{fig:shearfig}).
We attempted to do so using physically motivated perturbations from a simple ellipsoidal projected mass distribution, first with \(\Lambda\)CDM substructure (Figure \ref{fig:1cdmfsq}),
then with 10\(\times\Lambda\)CDM substructure (Figure \ref{fig:10xCDMpop}),
and later using a superposition of dark matter and baryon profiles with Fourier perturbations, misaligned PAs, and/or offset centers to alter
the shape of the mass isodensity contours (Figure \ref{fig:ellpert}). We devised a selection bias that mimics observational bias, based on the summed 
magnifications of the images, and applied it to our synthetic quads before comparing with observations.
Table \ref{table:summary} catalogs each experiment discussed herein and the corresponding p-values and relevant figures for each set of perturbations from purely 
elliptical lenses. One caveat to note is that the p-values quoted do not take into account the number of parameters for each model, so they cannot be properly 
compared to each other in any attempt to select a ``correct'' model, but only as a tool to judge which quad populations are consistent with the null hypothesis.

The first finding of this study is that substructure as predicted from \(\Lambda\)CDM is unable to generate quads with sufficient spread in \(\Delta\theta_{23}\) 
from the FSQ to match observations. Even if subhalos are ten times as massive as \(\Lambda\)CDM simulations predict, the resulting population of lenses 
is still insufficient to generate quads consistent with observations.
Factoring in external shear using realistic shear values from \citet{slacs5} does not alleviate this mismatch.

The second finding is that there exist some perturbations from pure ellipsoidal mass distributions which are capable of generating a population of quads consistent with observations, 
if an optimistic selection bias is applied. The most effective perturbation appears to be the one that models galaxies as two 
superimposed elliptical distributions, one representing dark matter and the other representing baryons, with the two centers offset by up to 1 kpc. Such a perturbation does not necessarily 
imply that the centers of baryon and dark matter distributions are 
not coincident, but rather introduces an asymmetry 
which causes a deviation from ellipticity near the image circle. When this perturbation is applied in conjunction 
with misaligning the PAs and applying realistic Fourier perturbations, the population consists of lenses like those in Figure \ref{fig:contourpopex}.
These galaxies bear visual resemblance to observed galaxies depicted in Figure 12 of \citet{mitsuda16}. After the bias is applied, the generated population of quads is consistent
with the observed population with a p-value of 6.2\%. 

It is interesting that these non-elliptical lens profiles create a better match with observations than elliptical ones.
This is not the first finding to indicate that this may be the case.
\citet{biggs04} studied a radio jet lensed by a galaxy in which spectral features made it possible to associate images with various features in the source. 
They found that a Singular Isothermal Ellipsoid with external shear was unable to account for the positions of all images simultaneously. The model they found which fit the 
image positions required drastic azimuthal dependence using a sum of Fourier coefficients, implying that the mass 
distribution for the galaxy had considerable ``wavy'' features. 

Why galaxies should be structured this way is an interesting question. 
For one reason or another, the dark matter mass distributions of these galaxies, though in equilibrium, are not relaxed \citep{Young16}, and additional 
complications from baryon-dark matter gravitational interactions, or dark matter self-interactions, are likely to make these systems even less relaxed.
Perhaps galaxy mergers are frequent and elliptical profiles are disrupted by such events. 
The answers to these questions will be vital to an understanding of galaxy formation. 

There remain other potential explanations which could cause the population of quads to not lie on the FSQ. We explored one in particular, albeit not at the level of detail with which we explored
substructure or superimposed but nonidentical baryon and dark matter profiles. This possibility is that of supermassive black holes (SMBHs) which are displaced from the center of the galaxy due 
to recoil from gravitational wave emission. 
These SMBHs are thought to be formed via merged black holes, which results in an asymmetry in the gravitational wave emission, imparting a recoil velocity on the SMBH which
can be on the order of the escape velocity, kicking the SMBH far from the center of the galaxy \citep{blecha16}. It has been predicted that if such systems exist, SMBHs would be present 
\(\sim1-10\) kpc away from the galactic center. To simulate this, we added a \(10^9\text{M}_{\odot}\) point mass to the synthetic lenses at a distance of 3 kpc, right at the image radius where it would
have the most effect on image positions.
Any alteration to the mass distribution was not visible on the density contour plot, and no substantial deviations from the FSQ were produced. While interesting, recoiling SMBHs are unlikely to 
have an effect on the quad population.

Another possible way for quads which lie off of the FSQ to be formed is for mass to be present between the source and the observer along the line of sight (LoS) \citep{mccully16}. Such structure would
make the thin-lens approximation used in many lens models inaccurate. Quads lensed with line of sight structure will be explored more thoroughly in a coming paper, although
the contribution of LoS and environmental structures near the main lens is unlikely
to be important. A number of papers have shown that LoS substructure contributes
less to the lensing optical depth than the substructure around the main lens
\citep{Met2002, chen2003, Wamb2005}\footnote{Note that \citet{li2017}, who conclude 
that LoS structures contribute more than the immediate environment of the main lens,
do not weigh their mass by the critical surface density for lensing, as is done in
other works.}, and if 10$\times\Lambda$CDM does not come close to observations in the
space of quad relative image angles, LoS is very unlikely to do so.

As this has been a preliminary study, there are several opportunities for future work. It is somewhat disconcerting that our synthesized population which
most closely matches observations has a relatively low p-value of only 6.2\%. We suspect that this is because the variables used to create non-ellipticity in the present paper
span a large parameter space which has not been fully explored. This may be a task suitable for machine learning, which could potentially identify new ways to 
create alterations from ellipticity in a way that 
results in a closer match between synthesized and observed quad populations. Additionally, as larger surveys come online, we can expect to find many more quad 
lenses. \citet{oguri10} predict that the Large Synoptic Survey Telescope will find \(\sim8000\) lensed quasars. Assuming a quad fraction similar to the 
present observed fraction, 0.154 \citep{oguri07}, one should anticipate over 1000 new quads to be discovered. 
Additionally, since all of these quads will be found in the same survey, the observational selection biases will 
be much easier to quantify and can be more reliably applied.
An analysis similar to ours using the larger sample size with well-determined biases will be more conclusive as to what type of galaxy lenses are consistent with observations.

\begin{sidewaystable*}[p]
\centering
\setlength\tabcolsep{6pt} 
\begin{tabular}{ |c|c|c|c|c|c|c|c|c|c|c|c|  }

 \multicolumn{1}{c}{}& \multicolumn{2}{|c|}{Dark matter only}& \multicolumn{2}{c}{}&\multicolumn{3}{|c|}{Baryons and dark matter} \\
 \cline{2-3} \cline{6-8}
 \hline 
 Population?& \(\Lambda\)CDM&\(10\times\Lambda\)CDM&Shear&Axis Ratios&Fourier&Misaligned Axes&Offset Centers&Bias?&p-value&FSQ Fig.&Mass Dist. Fig.\\
 \hline
   & x &   &   &   &   &   &   &   &\(2.6\times10^{-5}\%\)&Fig. \ref{fig:1cdmfsq}&Fig. \ref{fig:cdmcontour}\\
\hline
   &   & x &   &   &   &   &   &   &4.1\%&Fig. \ref{fig:sing10xcdmfsq} (middle)&Fig. \ref{fig:10xCDMcontour}\\
\hline
 x &   & x &   &   &   &   &   &   &\(0.099\%\)&Fig. \ref{fig:10xCDMpop}&Fig. \ref{fig:10xCDMcontour}\\
\hline
 x &   & x & x & x &   &   &   &   &\(0.0077\%\)& -- & -- \\
\hline
   &   & x &   &   &   &   &   & x &32\%&Fig. \ref{fig:sing10xcdmfsq} (right)&Fig. \ref{fig:10xCDMcontour}\\
\hline
 x &   & x &   &   &   &   &   & x &\(1.3\%\)& -- &Fig. \ref{fig:10xCDMcontour}\\
\hline
 x &   & x & x & x &   &   &   & x &1.3\%& -- & -- \\
\hline
 x &   &   &   & x & x &   &   &   &0.00044\%& -- & -- \\
\hline
 x &   &   &   & x & x &   &   & x &0.013\%& -- & -- \\
\hline
 x &   &   &   & x &   & x &   &   &0.0017\%& -- & -- \\
\hline
 x &   &   &   & x &   & x &   & x &0.020\%& -- & -- \\
\hline
 x &   &   &   & x &   & x & x &   &0.053\%& -- & -- \\
\hline
 x &   &   &   & x &   & x & x & x &3.4\%& -- & -- \\
\hline
 x &   &   &   & x & x & x & x &   &0.16\%& -- &Fig. \ref{fig:contourpopex}\\
\hline
 x &   &   &   & x & x & x & x & x &6.2\%&Fig. \ref{fig:ellpert}&Fig. \ref{fig:contourpopex}\\
\hline
\end{tabular}
\caption{A table summarizing the combinations of effects explored in this paper. Reading horizontally across describes each lens or
population of lenses explored, where an ``x`` denotes which effects were included on that experiment. Experiments are listed in order of appearance
within this paper, with the first being the single lens with only \(\Lambda\)CDM substructure and the last being the population of lenses with the 
combination of effects due to baryons listed in \ref{ssec:combo}. 
The effects listed are, in order, 
whether a population of lenses was used as opposed to a single lens, 
whether or not \(\Lambda\)CDM substructure was included, 
whether or not \(10\times\Lambda\)CDM substructure was included, 
whether external shears were drawn from \citet{slacs5} or left as zero, 
whether the axis ratios were drawn from \citet{slacs5} or all set to 0.82, 
whether or not a baryon population with nonzero \(a_4\) and \(a_6\) Fourier perturbations were included in addition to a dark matter component with similar Fourier perturbations,
whether or not the baryon population had a misaligned major axis with respect to the dark matter, 
whether or not the baryon population had an offset center from the dark matter, 
and whether or not the most optimistic bias was applied.
The remaining columns depict the p-values for each experiment as well as the figures where one can find the distribution of quads relative to the FSQ
and/or the mass distribution contours, where applicable. 
Caution is necessary when comparing p-values as there has been no accounting for the addition
of parameters, so the p-values are not directly comparable.}
\label{table:summary}
\end{sidewaystable*}






\bibliographystyle{mnras}
\bibliography{mnrasdraft3}








\bsp	
\label{lastpage}
\end{document}